\documentclass[twocolumn]{aastex62}

\usepackage{textcomp}
\usepackage[section]{placeins}
\usepackage{lineno}

\graphicspath{{./}{figures/}}

\received{\today}
\submitjournal{ApJ}

\shorttitle{Proton Anisotropy with {\it Fermi}-LAT}
\begin{document}

\title{A Search for Cosmic-ray Proton Anisotropy with the Fermi Large Area Telescope}
\author{M.~Ajello}
\affiliation{Department of Physics and Astronomy, Clemson University, Kinard Lab of Physics, Clemson, SC 29634-0978, USA}
\author{L.~Baldini}
\affiliation{Universit\`a di Pisa and Istituto Nazionale di Fisica Nucleare, Sezione di Pisa I-56127 Pisa, Italy}
\author{G.~Barbiellini}
\affiliation{Istituto Nazionale di Fisica Nucleare, Sezione di Trieste, I-34127 Trieste, Italy}
\affiliation{Dipartimento di Fisica, Universit\`a di Trieste, I-34127 Trieste, Italy}
\author{D.~Bastieri}
\affiliation{Istituto Nazionale di Fisica Nucleare, Sezione di Padova, I-35131 Padova, Italy}
\affiliation{Dipartimento di Fisica e Astronomia ``G. Galilei'', Universit\`a di Padova, I-35131 Padova, Italy}
\author{K.~Bechtol}
\affiliation{Department of Physics, University of Wisconsin-Madison, Madison, WI 53706, USA}
\author{R.~Bellazzini}
\affiliation{Istituto Nazionale di Fisica Nucleare, Sezione di Pisa, I-56127 Pisa, Italy}
\author{E.~Bissaldi}
\affiliation{Dipartimento di Fisica ``M. Merlin" dell'Universit\`a e del Politecnico di Bari, I-70126 Bari, Italy}
\affiliation{Istituto Nazionale di Fisica Nucleare, Sezione di Bari, I-70126 Bari, Italy}
\author{R.~D.~Blandford}
\affiliation{W. W. Hansen Experimental Physics Laboratory, Kavli Institute for Particle Astrophysics and Cosmology, Department of Physics and SLAC National Accelerator Laboratory, Stanford University, Stanford, CA 94305, USA}
\author{R.~Bonino}
\affiliation{Istituto Nazionale di Fisica Nucleare, Sezione di Torino, I-10125 Torino, Italy}
\affiliation{Dipartimento di Fisica, Universit\`a degli Studi di Torino, I-10125 Torino, Italy}
\author{E.~Bottacini}
\affiliation{W. W. Hansen Experimental Physics Laboratory, Kavli Institute for Particle Astrophysics and Cosmology, Department of Physics and SLAC National Accelerator Laboratory, Stanford University, Stanford, CA 94305, USA}
\affiliation{Department of Physics and Astronomy, University of Padova, Vicolo Osservatorio 3, I-35122 Padova, Italy}
\author{T.~J.~Brandt}
\affiliation{NASA Goddard Space Flight Center, Greenbelt, MD 20771, USA}
\author{P.~Bruel}
\affiliation{Laboratoire Leprince-Ringuet, \'Ecole polytechnique, CNRS/IN2P3, F-91128 Palaiseau, France}
\author{S.~Buson}
\affiliation{Institut f\"ur Theoretische Physik and Astrophysik, Universit\"at W\"urzburg, D-97074 W\"urzburg, Germany}
\author{R.~A.~Cameron}
\affiliation{W. W. Hansen Experimental Physics Laboratory, Kavli Institute for Particle Astrophysics and Cosmology, Department of Physics and SLAC National Accelerator Laboratory, Stanford University, Stanford, CA 94305, USA}
\author{R.~Caputo}
\affiliation{Center for Research and Exploration in Space Science and Technology (CRESST) and NASA Goddard Space Flight Center, Greenbelt, MD 20771, USA}
\author{E.~Cavazzuti}
\affiliation{Italian Space Agency, Via del Politecnico snc, 00133 Roma, Italy}
\author{S.~Chen}
\affiliation{Istituto Nazionale di Fisica Nucleare, Sezione di Padova, I-35131 Padova, Italy}
\affiliation{Department of Physics and Astronomy, University of Padova, Vicolo Osservatorio 3, I-35122 Padova, Italy}
\author{G.~Chiaro}
\affiliation{INAF-Istituto di Astrofisica Spaziale e Fisica Cosmica Milano, via E. Bassini 15, I-20133 Milano, Italy}
\author{S.~Ciprini}
\affiliation{Istituto Nazionale di Fisica Nucleare, Sezione di Roma ``Tor Vergata", I-00133 Roma, Italy}
\affiliation{Space Science Data Center - Agenzia Spaziale Italiana, Via del Politecnico, snc, I-00133, Roma, Italy}
\author{J.~Cohen-Tanugi}
\affiliation{Laboratoire Univers et Particules de Montpellier, Universit\'e Montpellier, CNRS/IN2P3, F-34095 Montpellier, France}
\author{D.~Costantin}
\affiliation{University of Padua, Department of Statistical Science, Via 8 Febbraio, 2, 35122 Padova}
\author{A.~Cuoco}
\affiliation{RWTH Aachen University, Institute for Theoretical Particle Physics and Cosmology, (TTK),, D-52056 Aachen, Germany}
\affiliation{Istituto Nazionale di Fisica Nucleare, Sezione di Torino, I-10125 Torino, Italy}
\author{S.~Cutini}
\affiliation{Istituto Nazionale di Fisica Nucleare, Sezione di Perugia, I-06123 Perugia, Italy}
\author{F.~D'Ammando}
\affiliation{INAF Istituto di Radioastronomia, I-40129 Bologna, Italy}
\affiliation{Dipartimento di Astronomia, Universit\`a di Bologna, I-40127 Bologna, Italy}
\author{P.~de~la~Torre~Luque}
\affiliation{Dipartimento di Fisica ``M. Merlin" dell'Universit\`a e del Politecnico di Bari, I-70126 Bari, Italy}
\author{F.~de~Palma}
\affiliation{Istituto Nazionale di Fisica Nucleare, Sezione di Torino, I-10125 Torino, Italy}
\author{A.~Desai}
\affiliation{Department of Physics and Astronomy, Clemson University, Kinard Lab of Physics, Clemson, SC 29634-0978, USA}
\author{S.~W.~Digel}
\affiliation{W. W. Hansen Experimental Physics Laboratory, Kavli Institute for Particle Astrophysics and Cosmology, Department of Physics and SLAC National Accelerator Laboratory, Stanford University, Stanford, CA 94305, USA}
\author{N.~Di~Lalla}
\affiliation{Universit\`a di Pisa and Istituto Nazionale di Fisica Nucleare, Sezione di Pisa I-56127 Pisa, Italy}
\author{L.~Di~Venere}
\affiliation{Dipartimento di Fisica ``M. Merlin" dell'Universit\`a e del Politecnico di Bari, I-70126 Bari, Italy}
\affiliation{Istituto Nazionale di Fisica Nucleare, Sezione di Bari, I-70126 Bari, Italy}
\author{A.~Dom\'inguez}
\affiliation{Grupo de Altas Energ\'ias, Universidad Complutense de Madrid, E-28040 Madrid, Spain}
\author{S.~J.~Fegan}
\affiliation{Laboratoire Leprince-Ringuet, \'Ecole polytechnique, CNRS/IN2P3, F-91128 Palaiseau, France}
\author{Y.~Fukazawa}
\affiliation{Department of Physical Sciences, Hiroshima University, Higashi-Hiroshima, Hiroshima 739-8526, Japan}
\author{S.~Funk}
\affiliation{Friedrich-Alexander-Universit\"at Erlangen-N\"urnberg, Erlangen Centre for Astroparticle Physics, Erwin-Rommel-Str. 1, 91058 Erlangen, Germany}
\author{P.~Fusco}
\affiliation{Dipartimento di Fisica ``M. Merlin" dell'Universit\`a e del Politecnico di Bari, I-70126 Bari, Italy}
\affiliation{Istituto Nazionale di Fisica Nucleare, Sezione di Bari, I-70126 Bari, Italy}
\author{F.~Gargano}
\affiliation{Istituto Nazionale di Fisica Nucleare, Sezione di Bari, I-70126 Bari, Italy}
\author{D.~Gasparrini}
\affiliation{Space Science Data Center - Agenzia Spaziale Italiana, Via del Politecnico, snc, I-00133, Roma, Italy}
\affiliation{Istituto Nazionale di Fisica Nucleare, Sezione di Perugia, I-06123 Perugia, Italy}
\author{N.~Giglietto}
\affiliation{Dipartimento di Fisica ``M. Merlin" dell'Universit\`a e del Politecnico di Bari, I-70126 Bari, Italy}
\affiliation{Istituto Nazionale di Fisica Nucleare, Sezione di Bari, I-70126 Bari, Italy}
\author{F.~Giordano}
\affiliation{Dipartimento di Fisica ``M. Merlin" dell'Universit\`a e del Politecnico di Bari, I-70126 Bari, Italy}
\affiliation{Istituto Nazionale di Fisica Nucleare, Sezione di Bari, I-70126 Bari, Italy}
\author{M.~Giroletti}
\affiliation{INAF Istituto di Radioastronomia, I-40129 Bologna, Italy}
\author{D.~Green}
\affiliation{Max-Planck-Institut f\"ur Physik, D-80805 M\"unchen, Germany}
\author{I.~A.~Grenier}
\affiliation{AIM, CEA, CNRS, Universit\'e Paris-Saclay, Universit\'e Paris Diderot, Sorbonne Paris Cit\'e, F-91191 Gif-sur-Yvette, France}
\author{S.~Guiriec}
\affiliation{The George Washington University, Department of Physics, 725 21st St, NW, Washington, DC 20052, USA}
\affiliation{NASA Goddard Space Flight Center, Greenbelt, MD 20771, USA}
\author{K.~Hayashi}
\affiliation{Department of Physics and Astrophysics, Nagoya University, Chikusa-ku Nagoya 464-8602, Japan}
\author{E.~Hays}
\affiliation{NASA Goddard Space Flight Center, Greenbelt, MD 20771, USA}
\author{J.W.~Hewitt}
\affiliation{University of North Florida, Department of Physics, 1 UNF Drive, Jacksonville, FL 32224 , USA}
\author{D.~Horan}
\affiliation{Laboratoire Leprince-Ringuet, \'Ecole polytechnique, CNRS/IN2P3, F-91128 Palaiseau, France}
\author{G.~J\'ohannesson}
\affiliation{Science Institute, University of Iceland, IS-107 Reykjavik, Iceland}
\affiliation{Nordita, Royal Institute of Technology and Stockholm University, Roslagstullsbacken 23, SE-106 91 Stockholm, Sweden}
\author{M.~Kuss}
\affiliation{Istituto Nazionale di Fisica Nucleare, Sezione di Pisa, I-56127 Pisa, Italy}
\author{L.~Latronico}
\affiliation{Istituto Nazionale di Fisica Nucleare, Sezione di Torino, I-10125 Torino, Italy}
\author{J.~Li}
\affiliation{Deutsches Elektronen Synchrotron DESY, D-15738 Zeuthen, Germany}
\author{I.~Liodakis}
\affiliation{W. W. Hansen Experimental Physics Laboratory, Kavli Institute for Particle Astrophysics and Cosmology, Department of Physics and SLAC National Accelerator Laboratory, Stanford University, Stanford, CA 94305, USA}
\author{F.~Longo}
\affiliation{Istituto Nazionale di Fisica Nucleare, Sezione di Trieste, I-34127 Trieste, Italy}
\affiliation{Dipartimento di Fisica, Universit\`a di Trieste, I-34127 Trieste, Italy}
\author{F.~Loparco}
\affiliation{Dipartimento di Fisica ``M. Merlin" dell'Universit\`a e del Politecnico di Bari, I-70126 Bari, Italy}
\affiliation{Istituto Nazionale di Fisica Nucleare, Sezione di Bari, I-70126 Bari, Italy}
\author{P.~Lubrano}
\affiliation{Istituto Nazionale di Fisica Nucleare, Sezione di Perugia, I-06123 Perugia, Italy}
\author{S.~Maldera}
\affiliation{Istituto Nazionale di Fisica Nucleare, Sezione di Torino, I-10125 Torino, Italy}
\author{A.~Manfreda}
\affiliation{Universit\`a di Pisa and Istituto Nazionale di Fisica Nucleare, Sezione di Pisa I-56127 Pisa, Italy}
\author{G.~Mart\'i-Devesa}
\affiliation{Institut f\"ur Astro- und Teilchenphysik and Institut f\"ur Theoretische Physik, Leopold-Franzens-Universit\"at Innsbruck, A-6020 Innsbruck, Austria}
\author{M.~N.~Mazziotta}
\affiliation{Istituto Nazionale di Fisica Nucleare, Sezione di Bari, I-70126 Bari, Italy}
\author{M.~Meehan}
\email{mrmeehan@wisc.edu}
\affiliation{Department of Physics, University of Wisconsin-Madison, Madison, WI 53706, USA}
\author{I.Mereu}
\affiliation{Dipartimento di Fisica, Universit\`a degli Studi di Perugia, I-06123 Perugia, Italy}
\author{M.~Meyer}
\affiliation{W. W. Hansen Experimental Physics Laboratory, Kavli Institute for Particle Astrophysics and Cosmology, Department of Physics and SLAC National Accelerator Laboratory, Stanford University, Stanford, CA 94305, USA}
\author{P.~F.~Michelson}
\affiliation{W. W. Hansen Experimental Physics Laboratory, Kavli Institute for Particle Astrophysics and Cosmology, Department of Physics and SLAC National Accelerator Laboratory, Stanford University, Stanford, CA 94305, USA}
\author{N.~Mirabal}
\affiliation{NASA Goddard Space Flight Center, Greenbelt, MD 20771, USA}
\affiliation{Department of Physics and Center for Space Sciences and Technology, University of Maryland Baltimore County, Baltimore, MD 21250, USA}
\author{W.~Mitthumsiri}
\affiliation{Department of Physics, Faculty of Science, Mahidol University, Bangkok 10400, Thailand}
\author{T.~Mizuno}
\affiliation{Hiroshima Astrophysical Science Center, Hiroshima University, Higashi-Hiroshima, Hiroshima 739-8526, Japan}
\author{A.~Morselli}
\affiliation{Istituto Nazionale di Fisica Nucleare, Sezione di Roma ``Tor Vergata", I-00133 Roma, Italy}
\author{M.~Negro}
\affiliation{Istituto Nazionale di Fisica Nucleare, Sezione di Torino, I-10125 Torino, Italy}
\affiliation{Dipartimento di Fisica, Universit\`a degli Studi di Torino, I-10125 Torino, Italy}
\author{E.~Nuss}
\affiliation{Laboratoire Univers et Particules de Montpellier, Universit\'e Montpellier, CNRS/IN2P3, F-34095 Montpellier, France}
\author{N.~Omodei}
\affiliation{W. W. Hansen Experimental Physics Laboratory, Kavli Institute for Particle Astrophysics and Cosmology, Department of Physics and SLAC National Accelerator Laboratory, Stanford University, Stanford, CA 94305, USA}
\author{M.~Orienti}
\affiliation{INAF Istituto di Radioastronomia, I-40129 Bologna, Italy}
\author{E.~Orlando}
\affiliation{W. W. Hansen Experimental Physics Laboratory, Kavli Institute for Particle Astrophysics and Cosmology, Department of Physics and SLAC National Accelerator Laboratory, Stanford University, Stanford, CA 94305, USA}
\author{V.~S.~Paliya}
\affiliation{Deutsches Elektronen Synchrotron DESY, D-15738 Zeuthen, Germany}
\author{D.~Paneque}
\affiliation{Max-Planck-Institut f\"ur Physik, D-80805 M\"unchen, Germany}
\author{M.~Persic}
\affiliation{Istituto Nazionale di Fisica Nucleare, Sezione di Trieste, I-34127 Trieste, Italy}
\affiliation{Osservatorio Astronomico di Trieste, Istituto Nazionale di Astrofisica, I-34143 Trieste, Italy}
\author{M.~Pesce-Rollins}
\affiliation{Istituto Nazionale di Fisica Nucleare, Sezione di Pisa, I-56127 Pisa, Italy}
\author{F.~Piron}
\affiliation{Laboratoire Univers et Particules de Montpellier, Universit\'e Montpellier, CNRS/IN2P3, F-34095 Montpellier, France}
\author{T.~A.~Porter}
\affiliation{W. W. Hansen Experimental Physics Laboratory, Kavli Institute for Particle Astrophysics and Cosmology, Department of Physics and SLAC National Accelerator Laboratory, Stanford University, Stanford, CA 94305, USA}
\author{G.~Principe}
\affiliation{Friedrich-Alexander-Universit\"at Erlangen-N\"urnberg, Erlangen Centre for Astroparticle Physics, Erwin-Rommel-Str. 1, 91058 Erlangen, Germany}
\author{S.~Rain\`o}
\affiliation{Dipartimento di Fisica ``M. Merlin" dell'Universit\`a e del Politecnico di Bari, I-70126 Bari, Italy}
\affiliation{Istituto Nazionale di Fisica Nucleare, Sezione di Bari, I-70126 Bari, Italy}
\author{R.~Rando}
\affiliation{Istituto Nazionale di Fisica Nucleare, Sezione di Padova, I-35131 Padova, Italy}
\affiliation{Dipartimento di Fisica e Astronomia ``G. Galilei'', Universit\`a di Padova, I-35131 Padova, Italy}
\author{M.~Razzano}
\affiliation{Istituto Nazionale di Fisica Nucleare, Sezione di Pisa, I-56127 Pisa, Italy}
\affiliation{Funded by contract FIRB-2012-RBFR12PM1F from the Italian Ministry of Education, University and Research (MIUR)}
\author{S.~Razzaque}
\affiliation{Department of Physics, University of Johannesburg, PO Box 524, Auckland Park 2006, South Africa}
\author{A.~Reimer}
\affiliation{Institut f\"ur Astro- und Teilchenphysik and Institut f\"ur Theoretische Physik, Leopold-Franzens-Universit\"at Innsbruck, A-6020 Innsbruck, Austria}
\affiliation{W. W. Hansen Experimental Physics Laboratory, Kavli Institute for Particle Astrophysics and Cosmology, Department of Physics and SLAC National Accelerator Laboratory, Stanford University, Stanford, CA 94305, USA}
\author{O.~Reimer}
\affiliation{Institut f\"ur Astro- und Teilchenphysik and Institut f\"ur Theoretische Physik, Leopold-Franzens-Universit\"at Innsbruck, A-6020 Innsbruck, Austria}
\affiliation{W. W. Hansen Experimental Physics Laboratory, Kavli Institute for Particle Astrophysics and Cosmology, Department of Physics and SLAC National Accelerator Laboratory, Stanford University, Stanford, CA 94305, USA}
\author{D.~Serini}
\affiliation{Dipartimento di Fisica ``M. Merlin" dell'Universit\`a e del Politecnico di Bari, I-70126 Bari, Italy}
\author{C.~Sgr\`o}
\affiliation{Istituto Nazionale di Fisica Nucleare, Sezione di Pisa, I-56127 Pisa, Italy}
\author{E.~J.~Siskind}
\affiliation{NYCB Real-Time Computing Inc., Lattingtown, NY 11560-1025, USA}
\author{G.~Spandre}
\affiliation{Istituto Nazionale di Fisica Nucleare, Sezione di Pisa, I-56127 Pisa, Italy}
\author{P.~Spinelli}
\affiliation{Dipartimento di Fisica ``M. Merlin" dell'Universit\`a e del Politecnico di Bari, I-70126 Bari, Italy}
\affiliation{Istituto Nazionale di Fisica Nucleare, Sezione di Bari, I-70126 Bari, Italy}
\author{D.~J.~Suson}
\affiliation{Purdue University Northwest, Hammond, IN 46323, USA}
\author{H.~Tajima}
\affiliation{Solar-Terrestrial Environment Laboratory, Nagoya University, Nagoya 464-8601, Japan}
\affiliation{W. W. Hansen Experimental Physics Laboratory, Kavli Institute for Particle Astrophysics and Cosmology, Department of Physics and SLAC National Accelerator Laboratory, Stanford University, Stanford, CA 94305, USA}
\author{J.~B.~Thayer}
\affiliation{W. W. Hansen Experimental Physics Laboratory, Kavli Institute for Particle Astrophysics and Cosmology, Department of Physics and SLAC National Accelerator Laboratory, Stanford University, Stanford, CA 94305, USA}
\author{D.~F.~Torres}
\affiliation{Institute of Space Sciences (CSICIEEC), Campus UAB, Carrer de Magrans s/n, E-08193 Barcelona, Spain}
\affiliation{Instituci\'o Catalana de Recerca i Estudis Avan\c{c}ats (ICREA), E-08010 Barcelona, Spain}
\author{E.~Troja}
\affiliation{NASA Goddard Space Flight Center, Greenbelt, MD 20771, USA}
\affiliation{Department of Astronomy, University of Maryland, College Park, MD 20742, USA}
\author{J.~Vandenbroucke}
\email{justin.vandenbroucke@wisc.edu}
\affiliation{Department of Physics, University of Wisconsin-Madison, Madison, WI 53706, USA}
\author{M.~Yassine}
\affiliation{Istituto Nazionale di Fisica Nucleare, Sezione di Trieste, I-34127 Trieste, Italy}
\affiliation{Dipartimento di Fisica, Universit\`a di Trieste, I-34127 Trieste, Italy}
\author{S.~Zimmer}
\affiliation{Institut f\"ur Astro- und Teilchenphysik and Institut f\"ur Theoretische Physik, Leopold-Franzens-Universit\"at Innsbruck, A-6020 Innsbruck, Austria}
\affiliation{University of Geneva, D\'epartement de physique nucl\'eaire et corpusculaire (DPNC), 24 quai Ernest-Ansermet, CH-1211 Gen\`eve 4, Switzerland}

\collaboration{The Fermi-LAT Collaboration}

%\correspondingauthor{Matthew Meehan}
%\correspondingauthor{Justin Vandenbroucke}
%\email{mrmeehan@wisc.edu, justin.vandenbroucke@wisc.edu}

\begin{abstract}
The {\it Fermi} Large Area Telescope (LAT) has amassed a large data set of primary cosmic-ray protons throughout its mission. The LAT's wide field of view and full-sky survey capabilities make it an excellent instrument for studying cosmic-ray anisotropy. As a space-based survey instrument, the LAT is sensitive to anisotropy in both right ascension and declination, while ground-based observations only measure the anisotropy in right ascension. We present the results of the first ever proton anisotropy search using {\it Fermi} LAT. The data set uses eight years of data and consists of approximately 179 million protons above 78 GeV, enabling it to probe dipole anisotropy below an amplitude of $10^{-3}$, resulting in the most stringent limits on the declination dependence of the dipole to date. We measure a dipole amplitude $\delta = 3.9\pm1.5 \times 10^{-4}$ with a p-value of 0.01 (pre-trials) for protons with a minimum energy of 78 GeV. We discuss various systematic effects that could give rise to a dipole excess and calculate upper limits on the dipole amplitude as a function of minimum energy. The 95\% CL upper limit on the dipole amplitude is $\delta_{UL}=1.3\times 10^{-3}$ for protons with a minimum energy of 78 GeV and $\delta_{UL}=1.2 \times 10^{-3}$ for protons with a minimum energy of 251 GeV.
\end{abstract}


\section{Introduction}\label{sec:intro}
Galactic cosmic rays diffuse through interstellar magnetic fields toward Earth, where they arrive with a high degree of isotropy. However, a small anisotropy in the arrival directions of cosmic rays of $\mathcal{O}(10^{-4}-10^{-3})$ has been consistently observed over the past several decades. The cosmic-ray anisotropy landscape has recently grown more complex as large experiments with long duty cycles have measured the anisotropy over nine decades in energy with unprecedented precision~\citep{amenomori_large-scale_2005,amenomori_anisotropy_2006,amenomori_northern_2017, abdo_discovery_2008, abdo_large-scale_2009, aglietta_evolution_2009, abbasi_measurement_2010, abbasi_observation_2011, abbasi_observation_2012, aartsen_observation_2013, aartsen_anisotropy_2016, bartoli_medium_2013, bartoli_argo-ybj_2015, abeysekara_observation_2014, abeysekara_observation_2018, abeysekara_all-sky_2018}. Broadly speaking, the anisotropy can be described by a large-scale, dipole-like feature with an energy-dependent amplitude and phase. Anisotropy at medium to smaller scales has also been measured down to angular scales of $\sim$10\textdegree, though with amplitudes an order of magnitude lower than the large-scale anisotropy~\citep{abbasi_measurement_2010, abeysekara_observation_2014, abeysekara_observation_2018}. A variety of physical mechanisms could explain the large-scale anisotropy, though there is no consensus on the exact causes of the energy dependence. Standard diffusion theory predicts a dipole in the direction of the density gradient of cosmic rays, but the predicted amplitude is up to two orders of magnitude larger than the observed anisotropy~\cite{hillas_can_2005, ptuskin_effect_2006, blasi_diffusive_2012}. The observed large-scale anisotropy at Earth could be due to the particular distribution of sources nearby as well as their directions relative to the local interstellar magnetic field~\citep{schwadron_consistent_2015, mertsch_solution_2015}. Observational effects of measuring the anisotropy with Earth-fixed observatories also obscure the true dipole. For example, the analysis techniques used to reach the appropriate sensitivity are incapable of measuring anisotropy along the declination axis, resulting in measurements along right ascension only. Partial sky coverage also biases the measurement of large-scale features. The all-sky anisotropy was recently measured using a combined data set from the IceCube and HAWC detectors, which demonstrated that previous measurements by either detector likely underestimate the large-scale amplitude~\citep{abeysekara_all-sky_2018}. It has been shown that the observed dipole can be explained by a combination of the astrophysical and instrumental effects described above~\citep{ahlers_deciphering_2016}. Many of the systematic effects introduced by the analysis techniques used by ground-based experiments can be mitigated by studying anisotropy with a full-sky, space-based observatory.

The {\it Fermi} Large Area Telescope (LAT) scans the entire celestial sky and detects cosmic rays in the GeV-TeV energy range. As a space-based survey instrument, it is sensitive to cosmic-ray anisotropy in both right ascension and declination. Additionally, the subsystems of the instrument can measure the charge of the cosmic rays, enabling a proton-only measurement of the anisotropy. The study of cosmic-ray anisotropy in this energy range offers complementary information to that at higher energies, as well as constraints on the declination dependence of the anisotropy which has never been measured. Furthermore, the study of anisotropy in this energy range is complementary to the study of the energy spectrum of protons and could shed light on the unexpected spectral break at $\sim$ a few hundred GeV~\citep{adriani_pamela_2011, aguilar_precision_2015}.

\section{Fermi Large Area Telescope}\label{sec:lat}
The LAT is a pair conversion gamma-ray telescope on board the {\it Fermi Gamma-ray Space Telescope} ({\it Fermi}) mission. Its wide field of view (2.4 sr) and full-sky survey capabilities make it an excellent instrument for studying cosmic-ray anisotropy. The {\it Fermi} spacecraft is in an equatorial orbit with an inclination of 25.6\textdegree. It rocks north and south from zenith toward the celestial poles on successive orbits, enabling the LAT to scan the entire sky every 2 orbits ($\sim$3 hours)~\footnote{The rocking angle of the instrument was 35\textdegree { }for the first year of data taking and increased to 50\textdegree   { }thereafter.}. The {\it Fermi}-LAT collaboration has published two studies of the anisotropy of cosmic-ray electrons and positrons, the latter of which provides the most stringent constraints on the dipole anisotropy of cosmic electrons and positrons to date~\citep{ackermann_searches_2010,abdollahi_search_2017}. The LAT has also amassed a large statistical sample of hadronic cosmic rays that can be studied using similar techniques to those in previous analyses. 

The LAT has three subsystems: an anti-coincidence detector (ACD) to reject the charged particle background, a tracker (TKR) to promote conversion of gamma rays to $e^+/e^-$ and measure their incident direction, and an electromagnetic calorimeter (CAL) to measure the energy of the resulting particle shower. The anti-coincidence detector consists of 89 segmented plastic scintillator panels covering the top and sides of the LAT. In cosmic-ray analyses, the ACD can be used to measure the charge of incident particles via their ionization losses through the scintillator tiles. The tracker is composed of 18 layers of x-y silicon strip detectors (SSDs) with interwoven tungsten foils to promote conversion of gamma rays into $e^+/e^-$ pairs. We used the tracker for direction reconstruction as well as a second, independent measurement of cosmic-ray charge. The calorimeter sits at the bottom of the LAT and consists of 1536 CsI(Tl) crystal logs in a hodoscopic arrangement, which allows for 3D reconstruction of the particle shower and is crucial for lepton-hadron separation. The imaging capabilities of the calorimeter also provide an independent, complementary direction measurement which we used to improve the angular resolution of the data set.

While the LAT's reconstruction algorithms are optimized for gamma rays, the instrument is essentially a charged lepton detector. The same basic principles for reconstructing particle direction and energy apply to both leptons and hadrons, though some care must be taken in understanding biases when measuring hadrons. The largest difference is in the energy estimate because the calorimeter is relatively shallow and does not fully contain hadronic showers. The calorimeter is 8.6~$X_0$ deep on axis, but only $\sim0.5$ hadronic interaction lengths. In general, the calorimeter contains the electromagnetic portion of the particle shower and underestimates the energy of hadronic showers. We apply a scaling relation developed with a Geant4~\citep{agostinelli_geant4simulation_2003} Monte Carlo simulation of protons interacting with the detector to account for the missing energy in the reconstruction and remove the bias when estimating proton energy.
  
\section{Data set}\label{sec:data_set}
In this section we describe the event selection and resulting data set used in the analysis. We selected eight years of data from 2008 December 26 to 2016 December 26 for this analysis. To ensure at least a minimum quality necessary to reconstruct events, a set of basic cuts was applied: we required that the events pass the onboard gamma filter~\footnote{At these energies, all showering cosmic rays should pass because any event with $>$20 GeV of raw energy, i.e., uncalibrated, deposited in the CAL passes the gamma filter.}, a track must be found, events must traverse a minimum of 4~$X_0$ of calorimeter integrated along their path length, and events must deposit a minimum of 21 GeV in the calorimeter~\footnote{The raw 20 GeV threshold used by the onboard filter is uncalibrated and affected by decreased light yields in the CAL crystals over time. The 21 GeV threshold on a calibrated quantity mitigates time dependence of the onbaord filter. Details of this effect are discussed in Section~\ref{sec:raw_thresh}.}. We also cut on a classification tree-based variable trained to identify accurately reconstructed events. These quality cuts are based on the event selection developed for the {\it Fermi}-LAT proton spectral measurement~\citep{green_measurement_2016, green_measurement_2017}. Additionally, we required that the LAT was in standard survey mode~\footnote{$\text{LAT\_MODE}$==5, $\text{LAT\_CONFIG}$==1, $\text{DATA\_QUAL}$==1, which are publicly available in the LAT pointing history (FT2) files.}, and that the rocking angle of the spacecraft was $<$52\textdegree~\footnote{This is to ensure that the Earth limb was well outside the field of view.}.

To separate protons from helium nuclei and other heavy cosmic rays we applied cuts on two independent measurements of the cosmic-ray charge, Z, using the tracker and the ACD. The signals in each subsystem are correlated with the charge due to the $Z^2$-scaling of the ionization loss. In the ACD, we measured the energy deposited in the single scintillator tile that is intersected by the best-fit track. A geometric correction was applied to account for the path length of each track. The average pulse height in the TKR provides a second, independent measurement of the charge. A two-dimensional cut on both variables was used to separate Z=1 particles from Z$>$1 particles. An additional cut on the energy deposited in ACD tiles within a 15\textdegree{ }cone of the best-fit track is also applied to remove residual nuclei, most of which have large incidence angles or enter the bottom of the detector. Residual contamination from helium and other Z$>$1 nuclei is estimated to be well below 1\%~\citep{green_measurement_2017}.

The charged-based selection described above yields a data set of protons and electrons. We used the electron classifier developed for the {\it Fermi}-LAT Pass 8 cosmic-ray electron and positron spectral and anisotropy analyses to separate protons from the remaining electrons~\citep{abdollahi_search_2017, abdollahi_cosmic-ray_2017}. The multivariate classifier uses the differences in the topology of electromagnetic and hadronic showers to separate the two event types. For example, the transverse width of hadronic showers is wider on average than that of electromagnetic showers. We refer to the most recent LAT CRE spectral measurement for more details on the classifier~\citep{abdollahi_cosmic-ray_2017}.

To mitigate the effects of the geomagnetic field, we imposed energy-dependent off-axis angle ($\theta$) cuts to reduce the LAT's field of view. This removes cosmic rays coming from the horizon, which have larger deflection angles than those with incident directions closer to zenith and are subjected to charge-dependent Earth shadowing. The maximum allowed off-axis angle is 45\textdegree{ }for protons with energies in (78 GeV, 139 GeV) and 50\textdegree{ }for all other events. Details of this selection are described in Section~\ref{sec:east_west}.

A detailed understanding of the instrument's point-spread function (PSF) is critical for a measurement of anisotropy. Compared to gamma rays, the LAT's angular resolution for protons is excellent: the 68\% containment is $\sim 0.01$\textdegree. However, Monte Carlo studies of the effect of the tails of the PSF on the measured anisotropy showed that events with large reconstruction errors can generate a false-positive signal. We utilized a second, independent measurement of the event direction from the 3D imaging calorimeter to reject poorly reconstructed events, primarily those that entered the bottom of the detector. We required that the angle between the best-fit track in the TKR and the best-fit direction in the CAL be $<0.2$ radians (11.5\textdegree). The procedure for determining this value without biasing the measurement is described in detail in Section~\ref{sec:psf_tail}.

\section{Anisotropy Search Method}\label{sec:method}

\subsection{Reference Map}\label{sec:reference}
To achieve the sensitivity necessary to measure anisotropy at these energies, the instrument's exposure cannot be calculated using simulation because the resulting uncertainties are significantly larger than the expected signal $\mathcal{O}(10^{-3})$. The observed data map is instead compared to a reference map, which is an estimate of the instrument's response to an isotropic cosmic-ray flux. Many data-driven methods of creating reference maps have been developed over the years to avoid relying on Monte Carlo simulation. The method we adopted uses the time-averaged event rate, $R_{avg}$, and distribution of detected event directions in instrument coordinates, $P(\theta,\phi)$, as empirical estimates of the detector's efficiency. In order to generate the reference map, the pointing history of the instrument was divided into 1 s time bins. For each 1 s bin with livetime $lt_{bin}$, the expected number of events was drawn from a Poisson distribution with mean $R_{avg} \times lt_{bin}$ and event directions were randomly drawn from $P(\theta,\phi)$. The sky directions for each event were then calculated for the position and orientation of the instrument at that time. To ensure that any anisotropic signal in the rate and $P(\theta,\phi)$ was adequately averaged out, we averaged over long time intervals of one year, i.e., $R_{avg}$ and $P(\theta,\phi)$ were calculated for each of the eight years in the data set. The choice of an integer number of years also minimizes contamination from the Compton-Getting dipole created by Earth's motion around the Sun, which cancels out in each complete year~\citep{compton_apparent_1935}. Additionally, the use of year-long time bins, rather than a single bin for the entire data set, mitigates the effects of decreasing light yields in the calorimeter which cause a small, monotonic decrease in the total event rate over the course of the LAT mission~\citep{bregeon_fermi-lat_2013}. One hundred independent reference maps were created and averaged to decrease the statistical uncertainty in the map. 

\begin{figure}[h]
	\centering
    \includegraphics[width=\columnwidth]{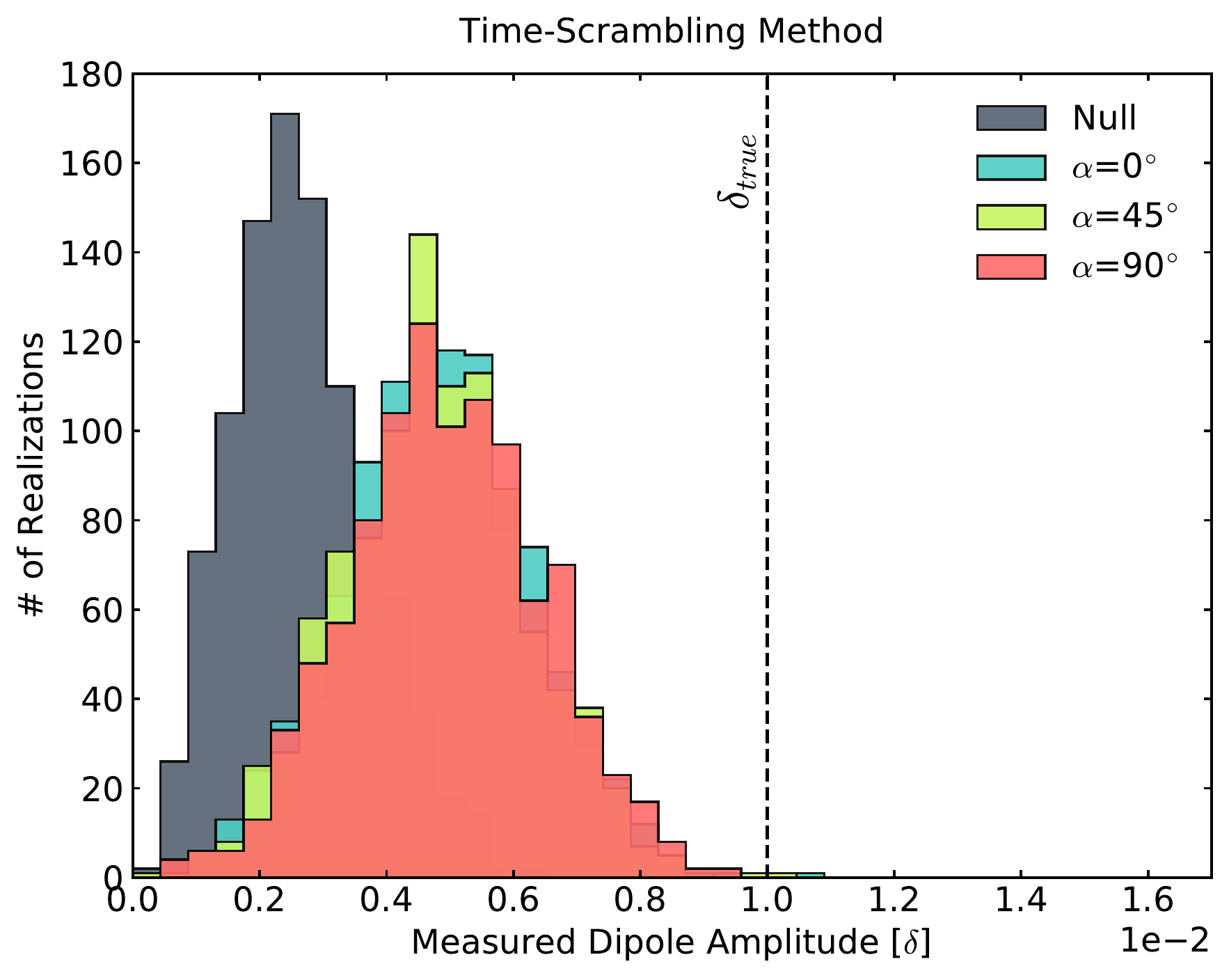}
    \includegraphics[width=\columnwidth]{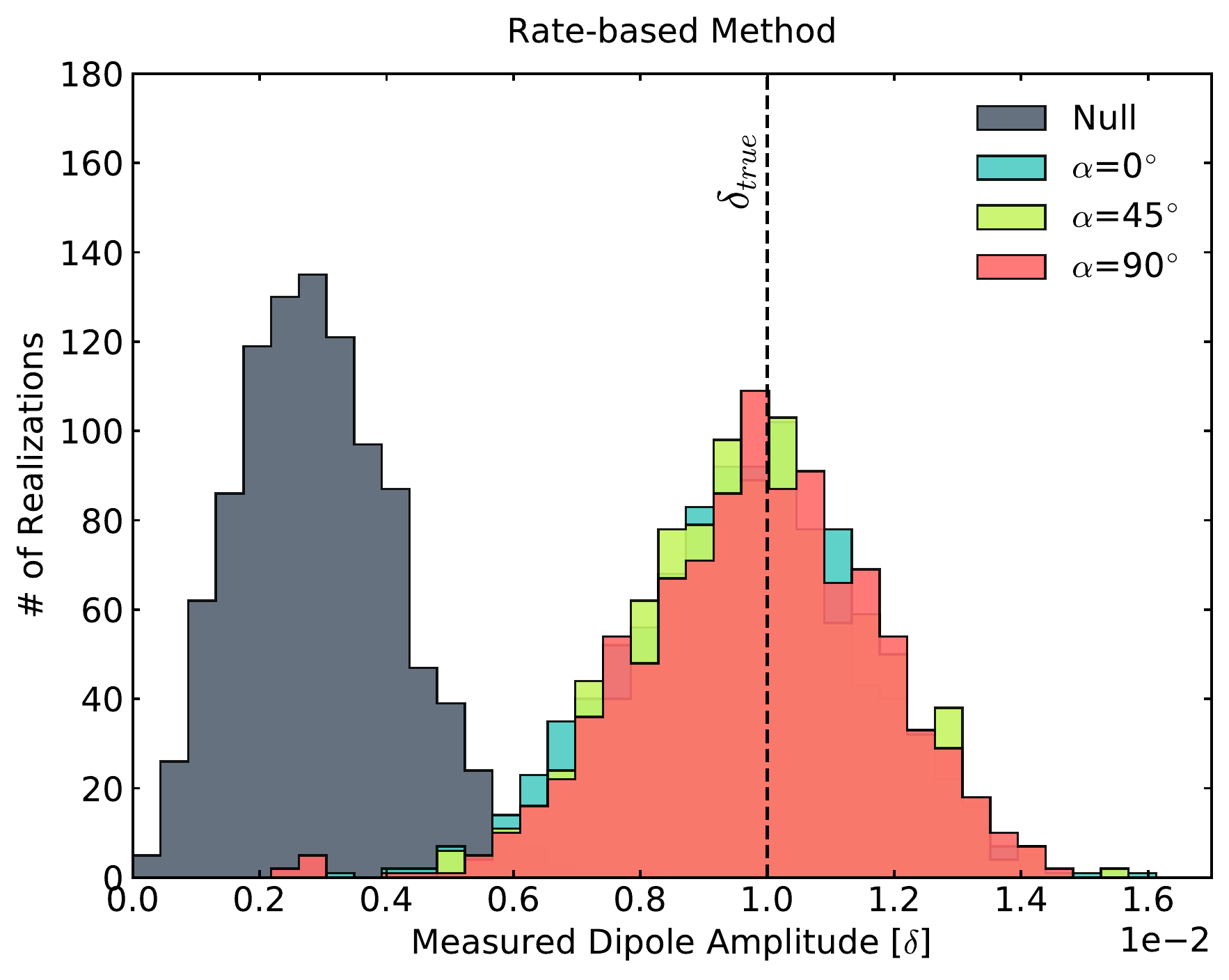}
	\caption{Results from an MC study comparing two reference map methods using an ideal detector with a conical field of view with a maximum off-axis angle of 60\textdegree. Each realization consists of one million events injected with a 1\% dipole oriented at an angle $\alpha$ relative to the declination axis. Reference maps were created using the time-scrambling method (top) and the rate-based method (bottom). The null hypothesis refers to realizations with events injected isotropically.}
	\label{fig:dipole_MC}
\end{figure}

We compared the rate-based reference map method to the time-scrambling~\citep{abbasi_measurement_2010, abbasi_observation_2011, abbasi_observation_2012, aartsen_observation_2013, aartsen_anisotropy_2016} or event-shuffling~\citep{ackermann_searches_2010,abdollahi_search_2017} commonly used for cosmic-ray anisotropy measurements. Using the time-scrambling method, the reference map is created by randomly shuffling the times of events in the data set and calculating new sky directions for each event with the shuffled time. We performed a Monte Carlo (MC) study to compare the performance of our rate-based method to the time-scrambling method when measuring a dipole using an ideal, conical detector with a 60\textdegree{ }opening angle. One million events were injected with a dipolar angular distribution with an amplitude of 0.01 and its maximum oriented at an angle $\alpha \in [0^{\circ}, 45^{\circ}, 90^{\circ}]$ relative to the polar axis. A null hypothesis data set was also produced by generating events isotropically. Reference maps were created using both the rate-based and time-scrambling methods and the dipole amplitude was reconstructed for each method. The results of one thousand realizations of this study are summarized in Figure~\ref{fig:dipole_MC}. The time-scrambling method consistently underestimates the true dipole amplitude by a factor of $\sim$2, while the rate-based method is unbiased. This is consistent with what was seen in the most recent {\it Fermi}-LAT anisotropy study of CREs, which compared four separate reference map techniques~\citep{abdollahi_search_2017}. A similar effect emerges with ground-based observatories at middle latitudes and iterative likelihood methods have been developed to unbias the measurement~\citep{ahlers_new_2016,abeysekara_observation_2018}. In both panels of Figure~\ref{fig:dipole_MC} the distribution of measured dipole amplitudes under the null hypothesis is centered around a non-zero value. This is the expected dipole amplitude due to Poissonian noise in the data set and determines the sensitivity of the analysis. Furthermore, neither method shows bias with respect to the direction of the dipole, i.e., the angle $\alpha$, indicating that both methods are sensitive to the right ascension and declination components of a dipole anisotropy. The rate-based reference map method is therefore sensitive to the two-dimensional direction of the dipole anisotropy.

\subsection{Spherical Harmonic Analysis}\label{sec:sph_harm_analysis}
After we generated the reference map, we compared it to the data map and performed a spherical harmonic analysis. Spherical harmonics offer a convenient way to characterize anisotropy at multiple angular scales. First, we calculated the relative intensity between the data map and the reference map:

\begin{equation}\label{eq:relint}
	\delta I_i(\theta,\phi) = \frac{N_i}{\left\langle N \right\rangle_i} - 1
\end{equation}
where $N_i$ and $\left\langle N \right\rangle_i$ are the counts in the $i$th pixel of the data map and reference map, respectively.

The relative intensity was then decomposed into spherical harmonics:

\begin{equation}\label{eq:sph_harm_expansion}
	\delta I(\theta,\phi) = \sum_{\ell=0}^\infty \sum_{m=-\ell}^\ell a_{\ell m} Y_{\ell m}(\theta,\phi)
\end{equation}
The coefficients of the expansion were computed using the anafast algorithm in the HEALPix~\footnote{http://healpix.sf.net} library~\citep{gorski_healpix:_2005}. The coefficients were then converted to coefficients of the real-valued, or tesseral, spherical harmonic functions for a more natural geometric interpretation. All of the sky maps were binned using the HEALPix scheme with an Nside parameter of 16 (3072 pixels), which corresponds to an angular extent of approximately 4 $\text{deg}$.  

The coefficients of the spherical harmonic decomposition can be used to characterize any anisotropy at each angular scale. The angular power at each multipole is calculated directly from the coefficients of the multipole expansion:
\begin{equation}\label{eq:power_spectrum}
	C_\ell = \frac{1}{2\ell+1}\sum_{m=-\ell}^\ell{\left\vert a_{\ell m} \right\vert^2}
\end{equation}

The measured angular power $C_\ell$ contains contributions from two terms: the true anisotropic power in the map $C_\ell^{true}$, which we are interested in measuring, and a noise term from randomly correlated statistical fluctuations in the map, $C_N$. $C_N$, which is equivalent to the variance of $a_{lm}$, can be calculated by propagating errors in the relative intensity map to the final quantities~\citep{knox_determination_1995}. The following expression accounts for pixel-to-pixel variation introduced by non-uniform exposure in the maps~\citep{fornasa_angular_2016, abdollahi_search_2017}:
\begin{equation}\label{eq:c_noise}
	C_N = \frac{4\pi}{N_{pix}^2} \sum_{i=1}^{N_{pix}} \frac{N_i}{\left\langle N_i \right\rangle ^2} + \alpha \frac{N_i^2}{\left\langle N_i \right\rangle ^3}
\end{equation}
where $N_{pix}$ is the number of pixels in the map, $\alpha = 1/n$, and $n$ is the number of reference maps created and averaged to estimate the isotropic expectation, which is 100 in this analysis. Note that in the case of uniform exposure and very large $n$, the formula simplifies to $C_N = \frac{4\pi}{N}$, where N is the total number of events in the map~\footnote{Uniform exposure refers to the case where $\left\langle N_i \right\rangle = \left\langle N_j \right\rangle \text{ for all } i,j$. In the large statistics limit, i.e., $N_i = \left\langle N_i \right\rangle$ and $\left\langle N_i \right\rangle = N/N_{pix}$, and with uniform exposure, Equation~\ref{eq:c_noise} results in this expression.}. To estimate the true anisotropic power in the map, we subtract the noise contribution, which results in the maximum likelihood estimator for the true anisotropy:

\begin{equation}\label{eq:c_noise}
	\hat{C_\ell}^{true} = C_\ell - C_N
\end{equation}

The angular power spectrum characterizes the total anisotropy at each angular scale, with angular features at each multipole $\ell$ $\sim$180\textdegree/$\ell$. Any excess or deficit in the angular power spectrum compared to the isotropic expectation indicates anisotropy at that angular scale. The expected distribution of angular power at each multipole under the null hypothesis can be calculated by assuming that $C_\ell$ follows a $\chi^2_{2\ell+1}$ distribution with mean $C_N$~\citep{knox_determination_1995}.

As described in the introduction, the dipole anisotropy is especially interesting scientifically. The amplitude of the dipole can be calculated directly from the angular power at multipole $\ell$=1:

\begin{equation}\label{eq:delta}
\delta = 3\sqrt{\frac{C_1}{4\pi}}
\end{equation}
and the estimator for the true dipole amplitude, $\hat{\delta}$, is calculated by inputting $\hat{C_1}^{true}$ into the above equation and imposing the requirement that $\hat{C_1}^{true}>0$. 

In addition to the total dipole amplitude, we can also calculate the full two-dimensional direction from the spherical harmonic coefficients. The right ascension of the dipole is given by:

\begin{equation}
\text{RA} = \arctan(\frac{a_{1-1}}{a_{11}})
\end{equation}
while the declination is given by:

\begin{equation}
\text{Dec} = \frac{\pi}{2} - \arccos(\sqrt{\frac{3}{4\pi}}\frac{a_{10}}{\hat{\delta}})
\end{equation}

The accompanying uncertainty in each quantity is calculated by propagating the statistical uncertainty in the relative intensity map to the derived quantities.

\section{Results}\label{sec:results}
\begin{figure*}[h]
	\centering
    \includegraphics[width=\columnwidth]{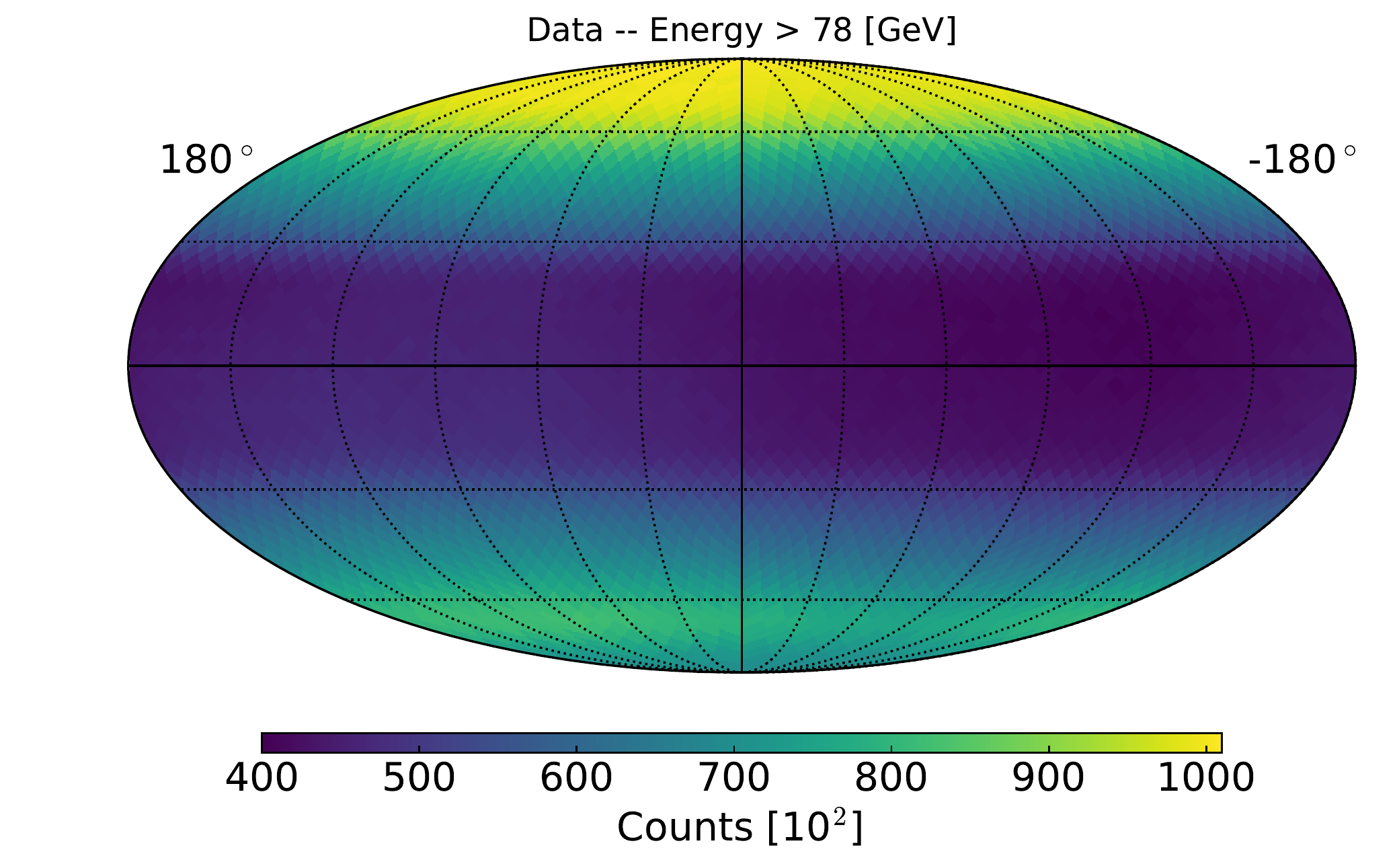}
    \includegraphics[width=\columnwidth]{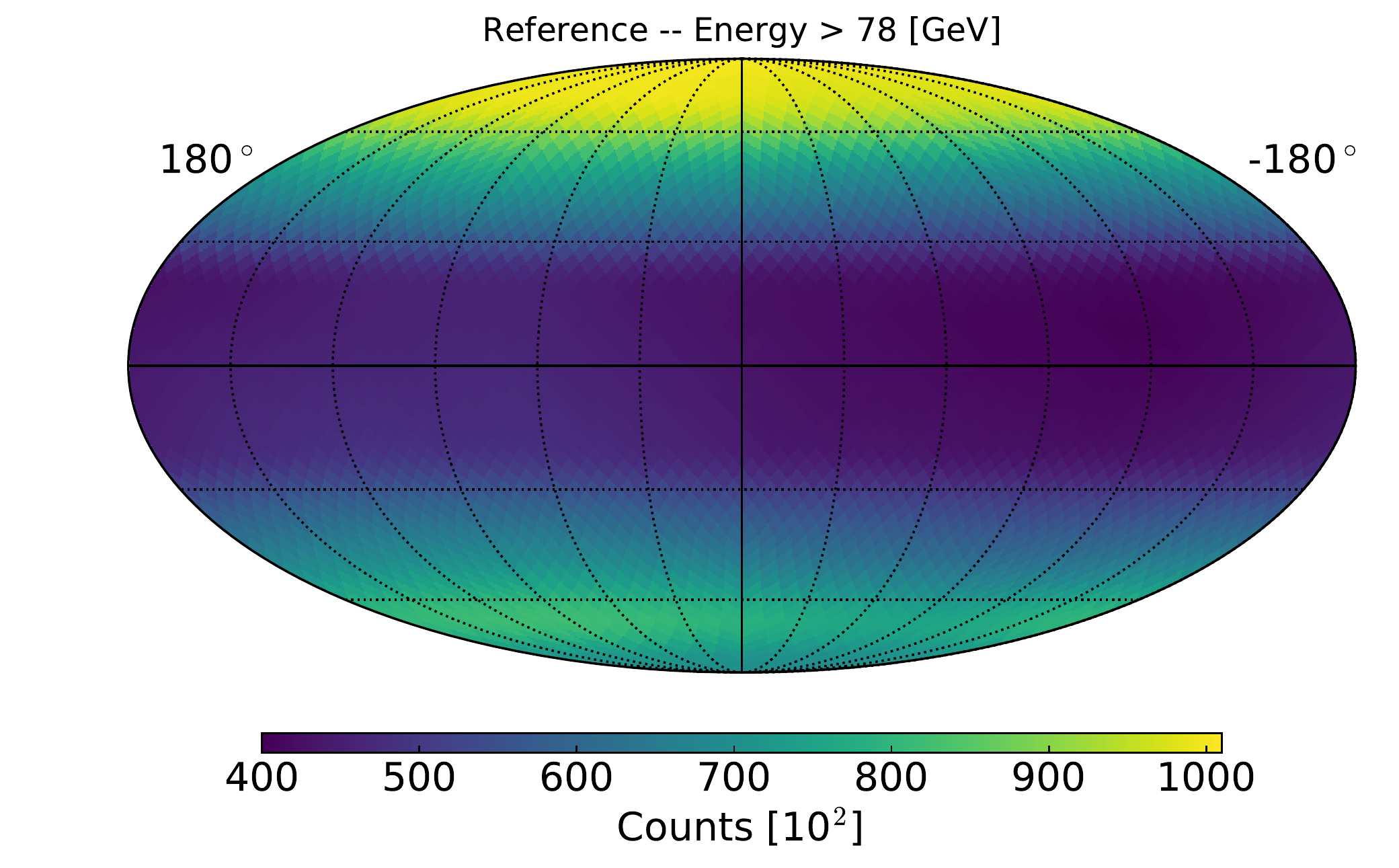}
	\caption{Data and reference sky maps in equatorial coordinates (J2000) for all events in the data set. Sky maps are created using the HEALPix pixelization scheme with 3072 pixels and use the astronomical convention of right ascension increasing to the left.}
	\label{fig:sky_maps}
\end{figure*}

\begin{figure*}[h]
	\centering
    \includegraphics[width=\columnwidth]{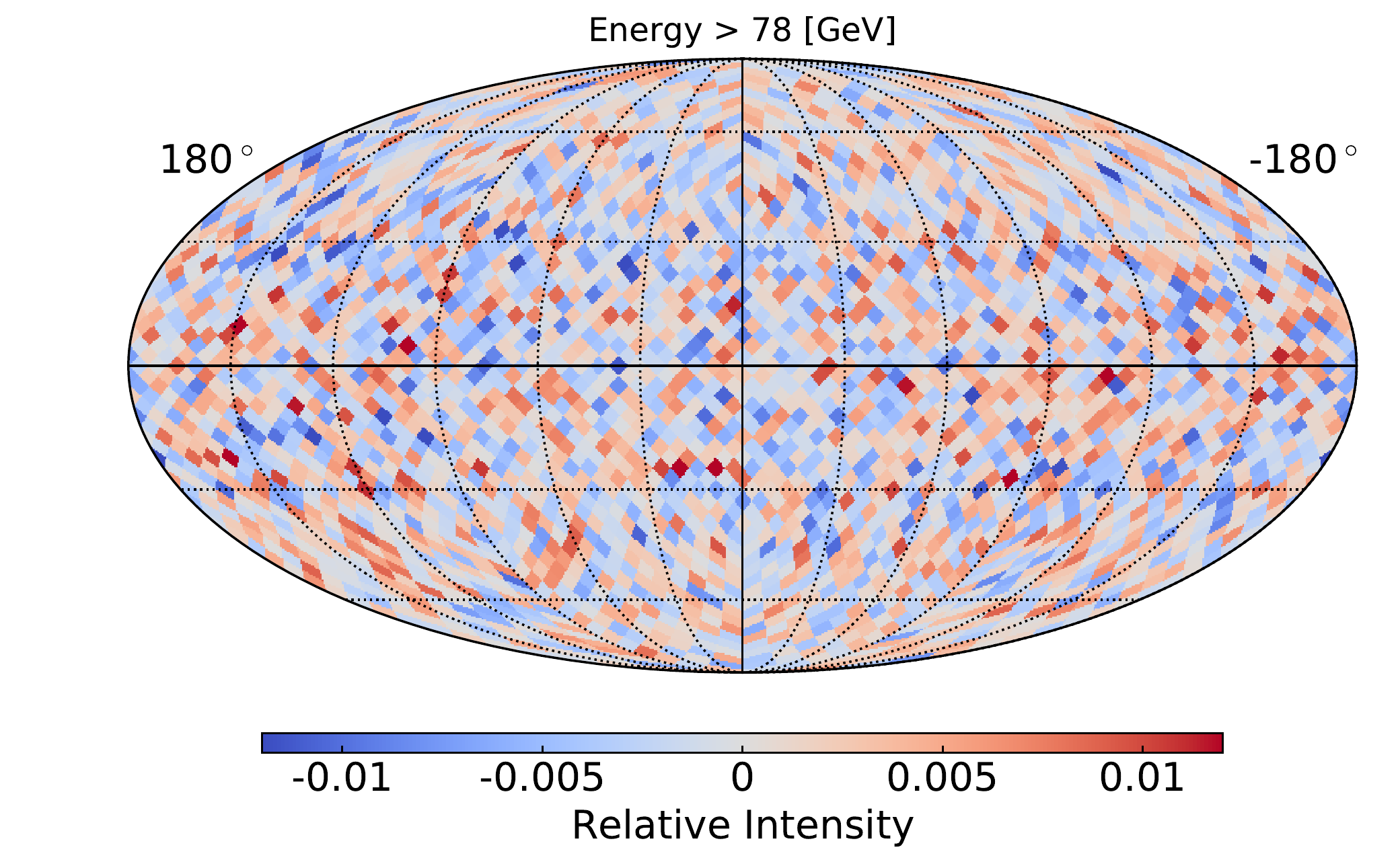}
    \includegraphics[width=\columnwidth]{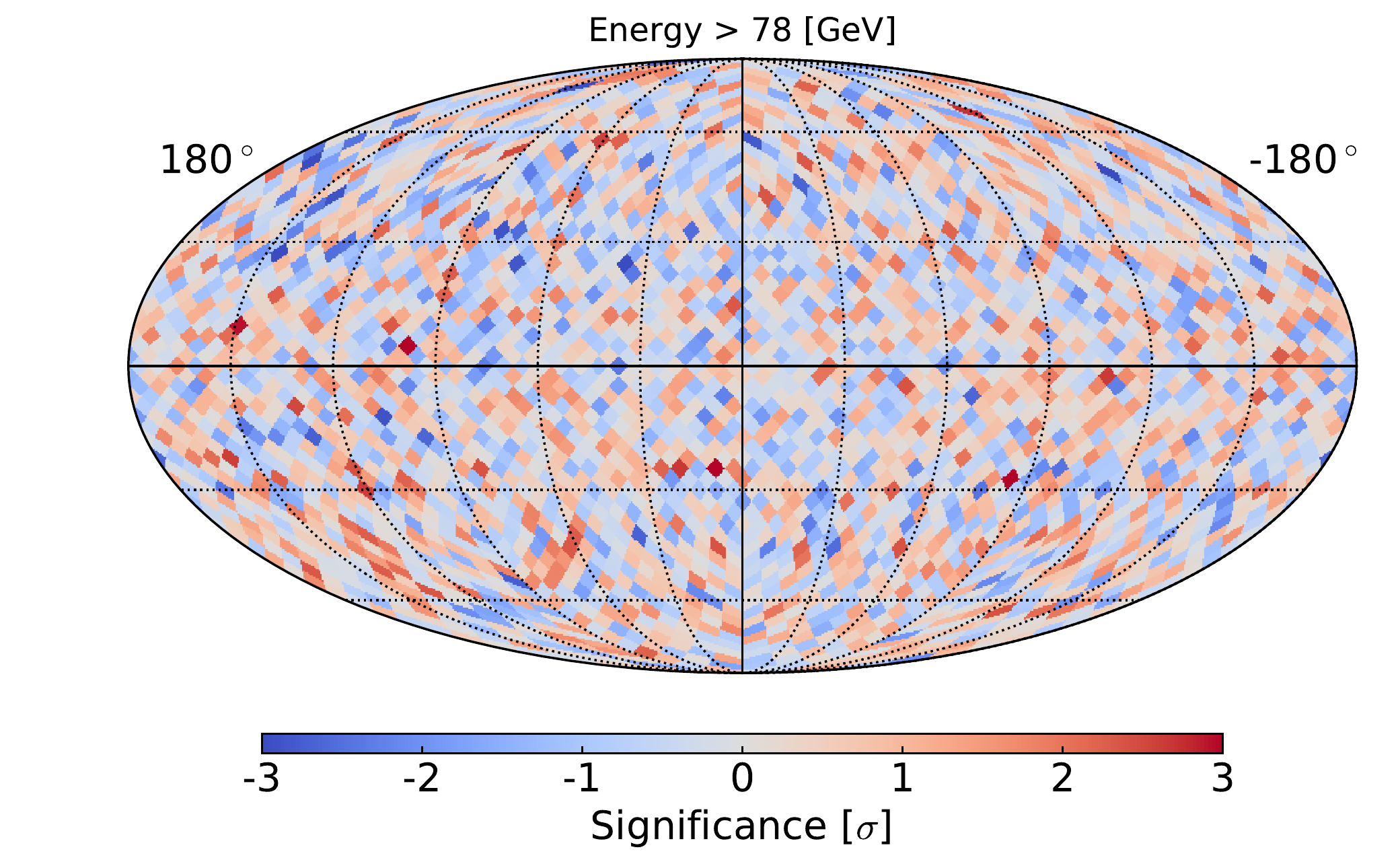}
	\caption{Relative intensity and significance maps in equatorial coordinates (J2000) for all events in the data set. Statistical fluctuations are smaller towards the equatorial poles because the exposure is greater towards the poles. Sky maps are created using the HEALPix pixelization scheme with 3072 pixels and use the astronomical convention of right ascension increasing to the left.}
	\label{fig:relint_maps}
\end{figure*}

In this section we present the results of the analysis performed using the methods described in the previous section. The data were divided into eight logarithmically-spaced energy bins in estimated gamma-ray energy which correspond to approximately 80 GeV - 10 TeV in estimated proton energy. Reference maps were generated for each energy bin using an averaging period, or time bin, of one year as described in Section~\ref{sec:reference}. The data maps and reference maps were summed over all eight years of the data set, and summed cumulatively in energy to maximize the sensitivity of the analysis. The data map and reference map for the minimum cumulative energy bin, i.e., the energy bin spanning the full data set, can be seen in Figure~\ref{fig:sky_maps}. The structure seen in both the data and reference maps is a result of the LAT's exposure, which is biased towards the northern and southern celestial poles due to the instrument's rocking profile. The relative intensity map, which is used for the spherical harmonic analysis, and a significance map are shown in Figure~\ref{fig:relint_maps}.

\begin{figure*}[h]
	\centering
    \includegraphics[width=0.3\linewidth]{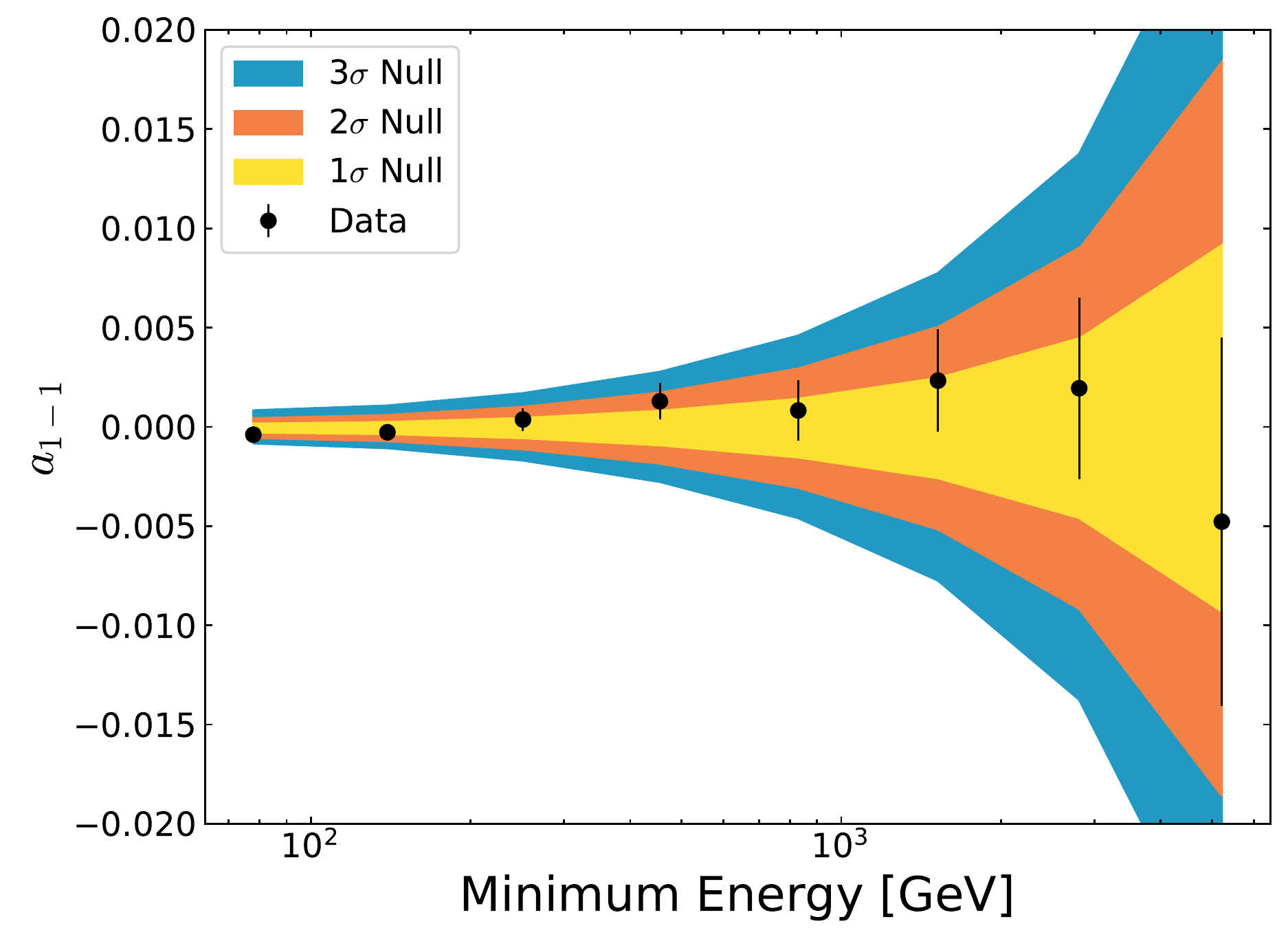}
    \includegraphics[width=0.3\linewidth]{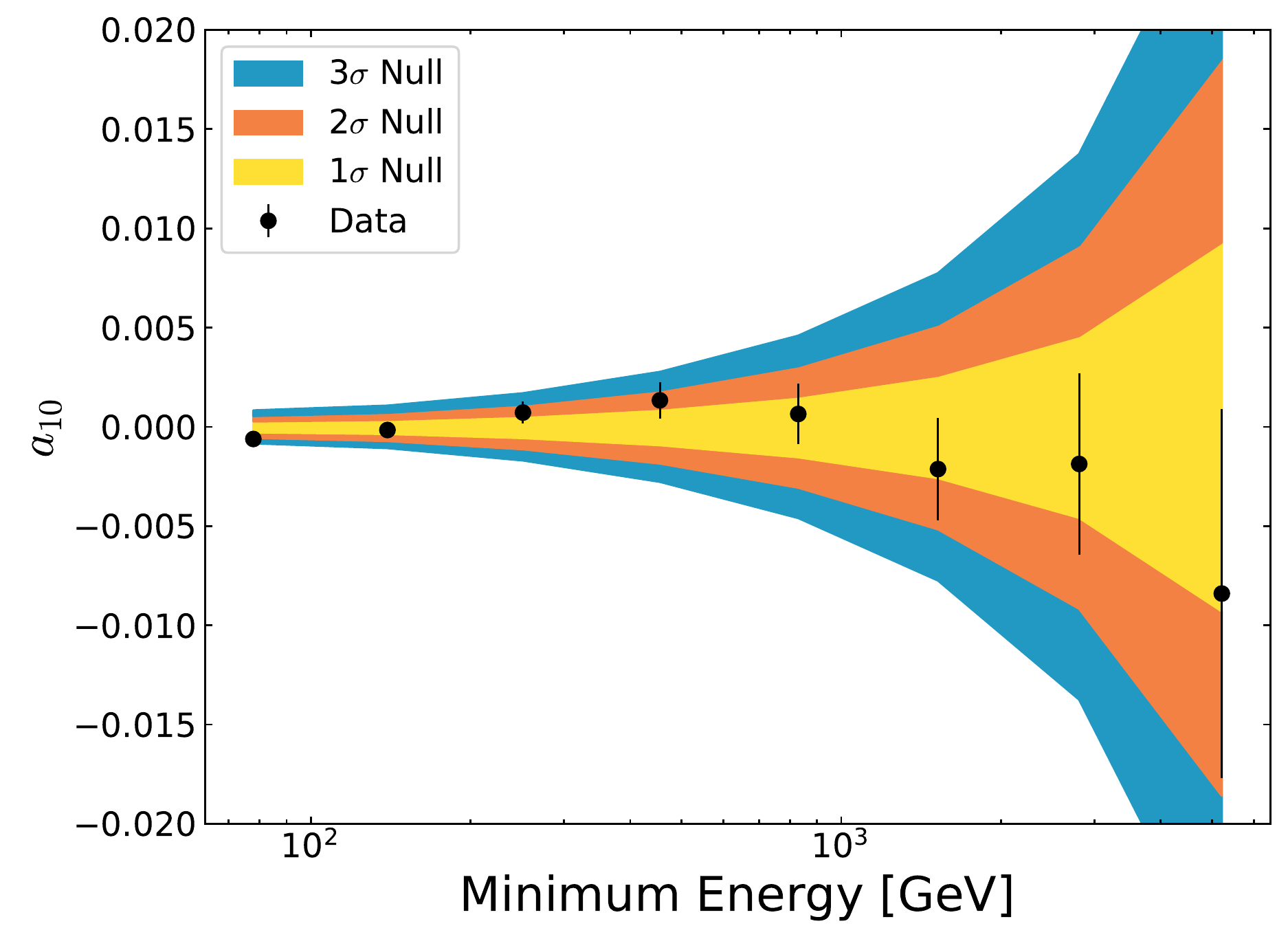}
    \includegraphics[width=0.3\linewidth]{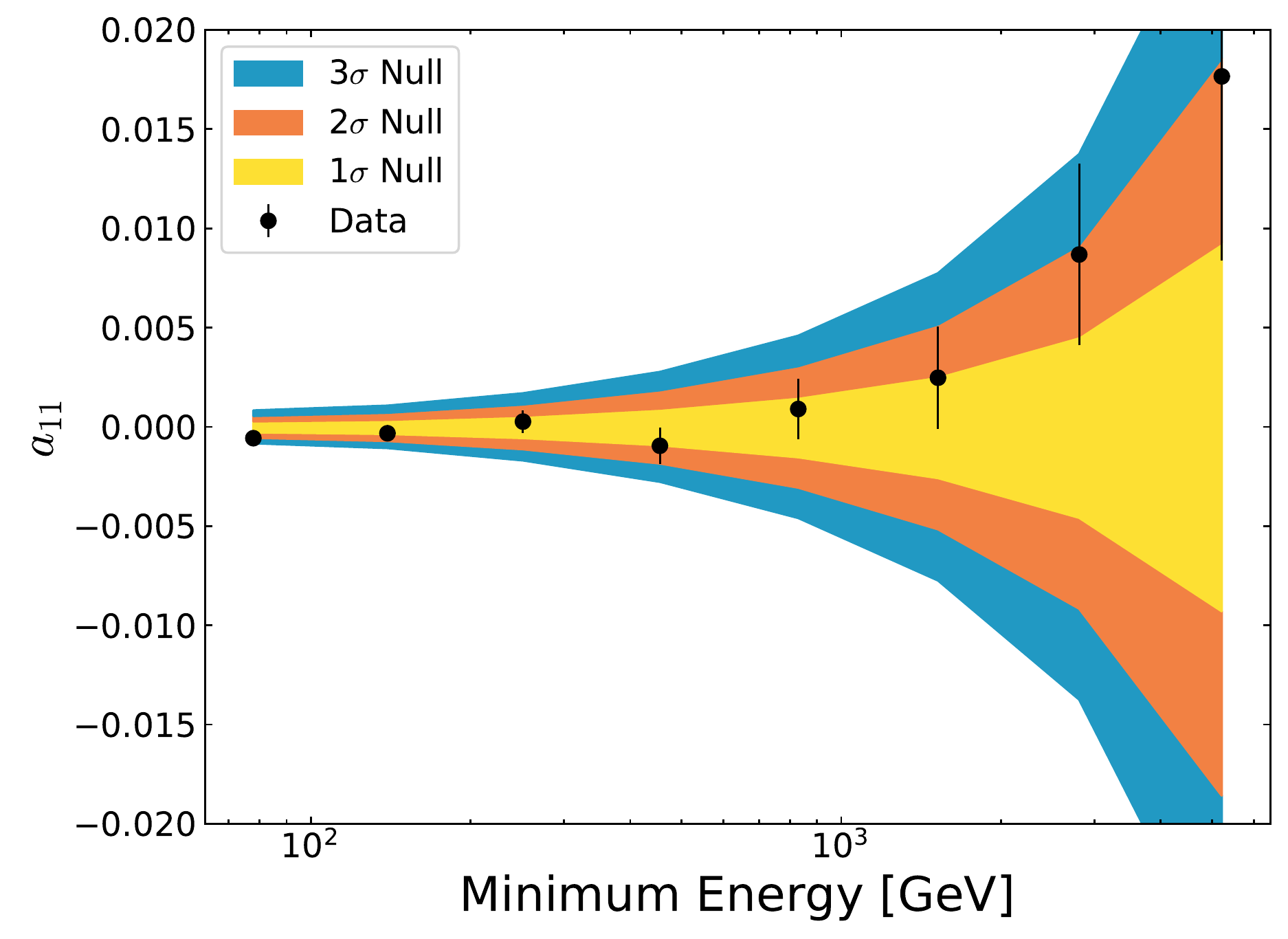}
	\caption{Measured $a_{1m}$ coefficients of the spherical harmonic expansion vs. minimum energy. From left to right: $a_{1-1}$, $a_{10}$, and $a_{11}$. Note that the $a_{10}$ coefficient corresponds to the dipole component aligned with the polar axis, which is unconstrained by ground-based measurements. The error bars are the statistical uncertainty on the measured coefficients and the colored bands represent the distribution of expected results under the null hypothesis, i.e., isotropic sky.}
	\label{fig:a1m_vs_energy}
\end{figure*}

\begin{figure}[h]
	\centering
    \includegraphics[width=\columnwidth]{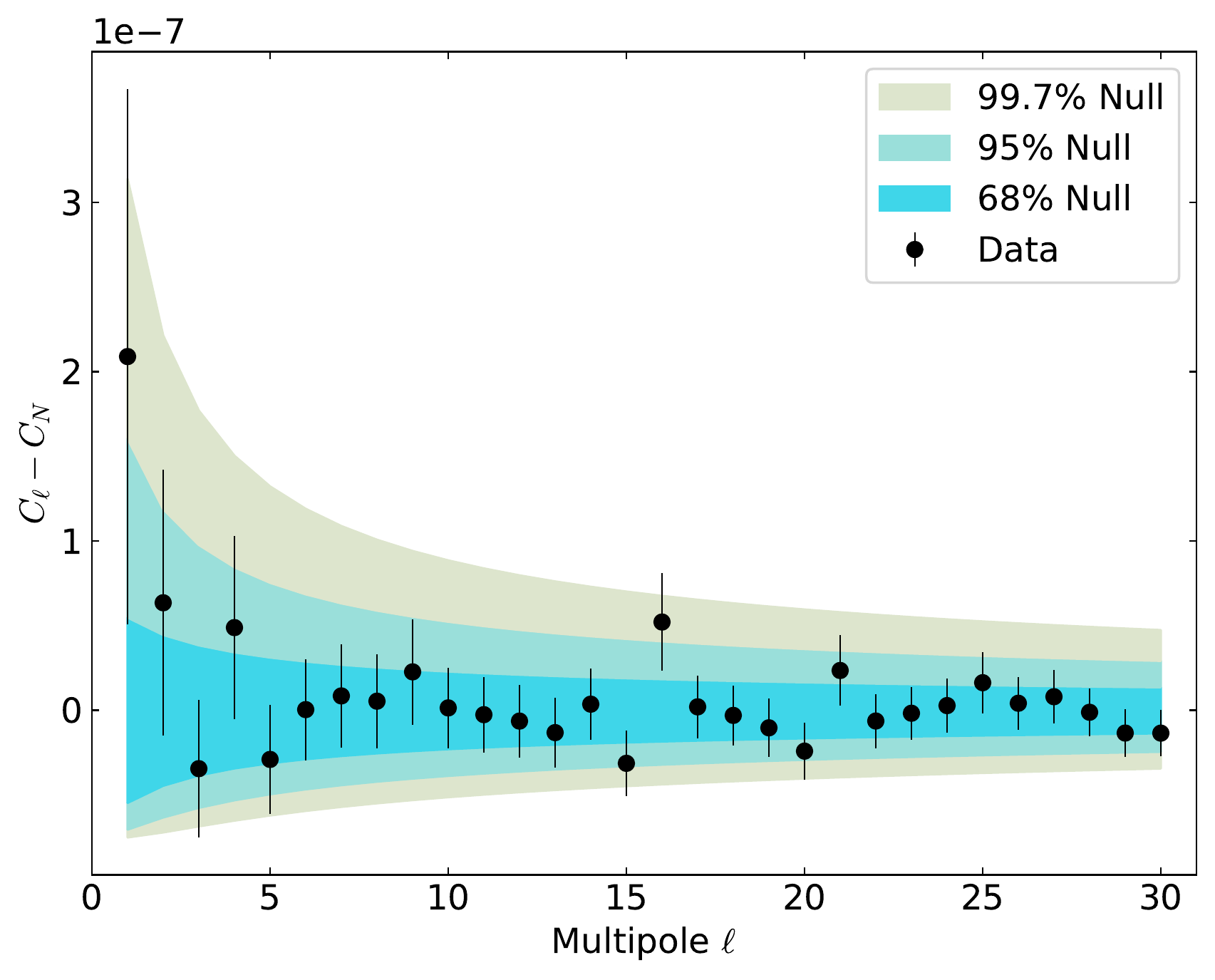}
	\caption{Angular power spectrum calculated for the relative intensity map with a minimum energy of 78 GeV. The horizontal axis is the multipole $l$ of the spherical harmonic expansion and the vertical axis is angular power at that $l$ value. The angular scale of each multipole is $\sim$180\textdegree/$\ell$. $C_\ell$ is the angular power in the map, which includes an anisotropic component and a noise component, $C_N$. The angular power due to white noise is subtracted from the measured power, which is the maximum likelihood estimator for the underlying anisotropy in the map. The error bars are the statistical uncertainty on the measured angular power. The colored bands represent the distribution of expected results under the null hypothesis, i.e., isotropic sky.}
	\label{fig:power_spectrum}
\end{figure}

We performed the spherical harmonic analysis detailed in Section~\ref{sec:sph_harm_analysis} on the relative intensity maps corresponding to each of the cumulative energy bins. The measured $a_{\ell m}$ coefficients, which contain the direction of the large-scale anisotropy, are plotted in Figure~\ref{fig:a1m_vs_energy} and their measured values are in Table~\ref{tbl:sph_harm_coeff}. The angular power spectrum for the entire energy range, i.e., E$>$78 GeV, is shown in Figure~\ref{fig:power_spectrum}. We calculated the angular power spectrum up to $\ell$=30, which corresponds to an angular scale of $\sim$6\textdegree. There is a weak excess in the dipole power, $C_1$, with a p-value of 0.01 (pre-trials). There is an additional excess at $\ell$=16. However, this is likely to be a statistical fluctuation. Each of the 30 angular power measurements is independent. Under the null hypothesis, the expectation value for the number of measurements that should exceed the 95\% interval due to random chance is 1.5. We note that either excess could be a statistical fluctuation, however, there is a known astrophysical anisotropy at $\ell=1$ and there are systematics that could potentially create a false-positive dipole excess that will be discussed in Section~\ref{sec:systematics}. The interpretation of the $\ell=1$, i.e., dipole, measurement is therefore more nuanced than that at $\ell=16$ and will be discussed in detail shortly.

\begin{figure}[h]
	\centering
    \includegraphics[width=\columnwidth]{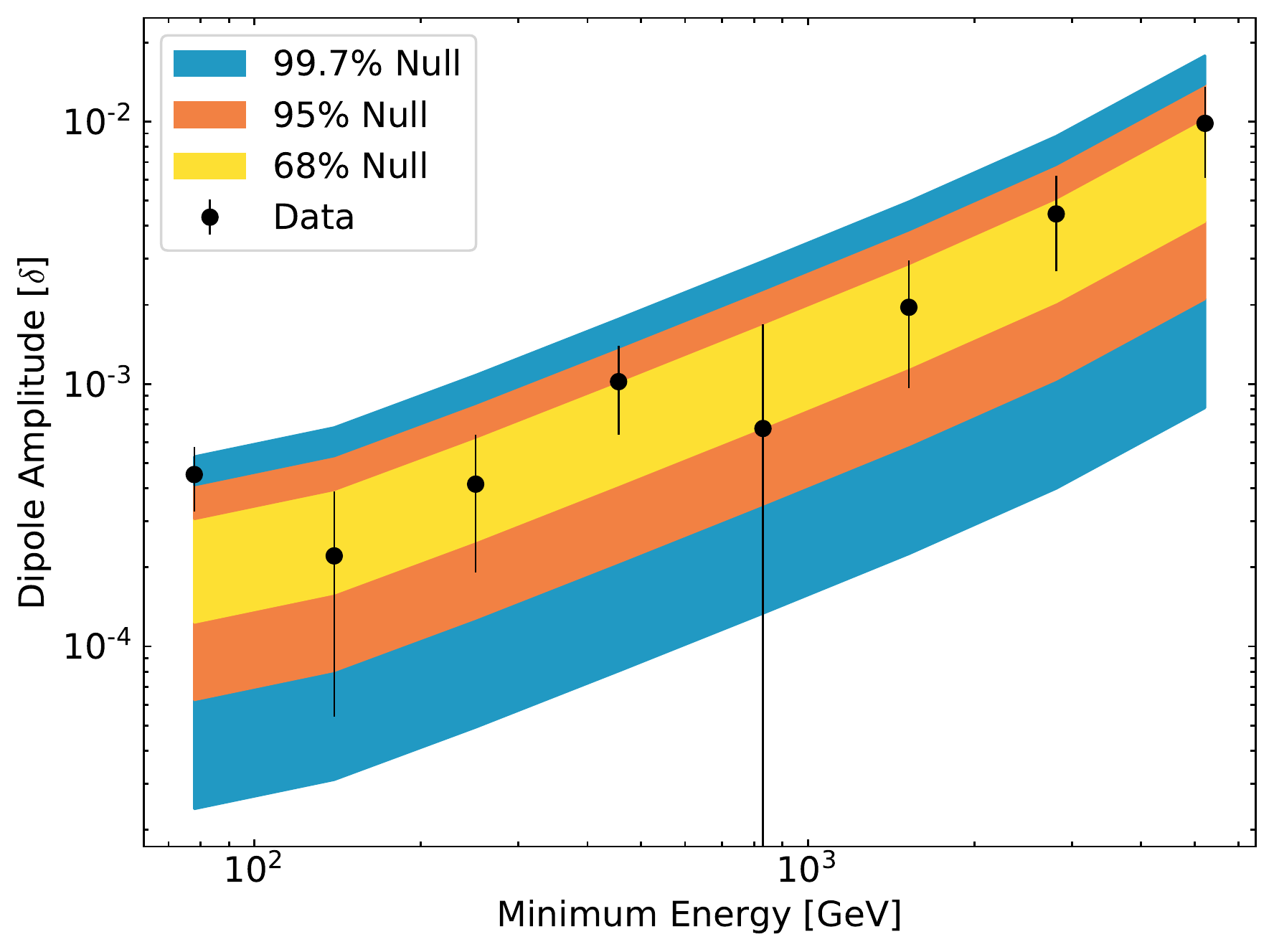}
    \includegraphics[width=\columnwidth]{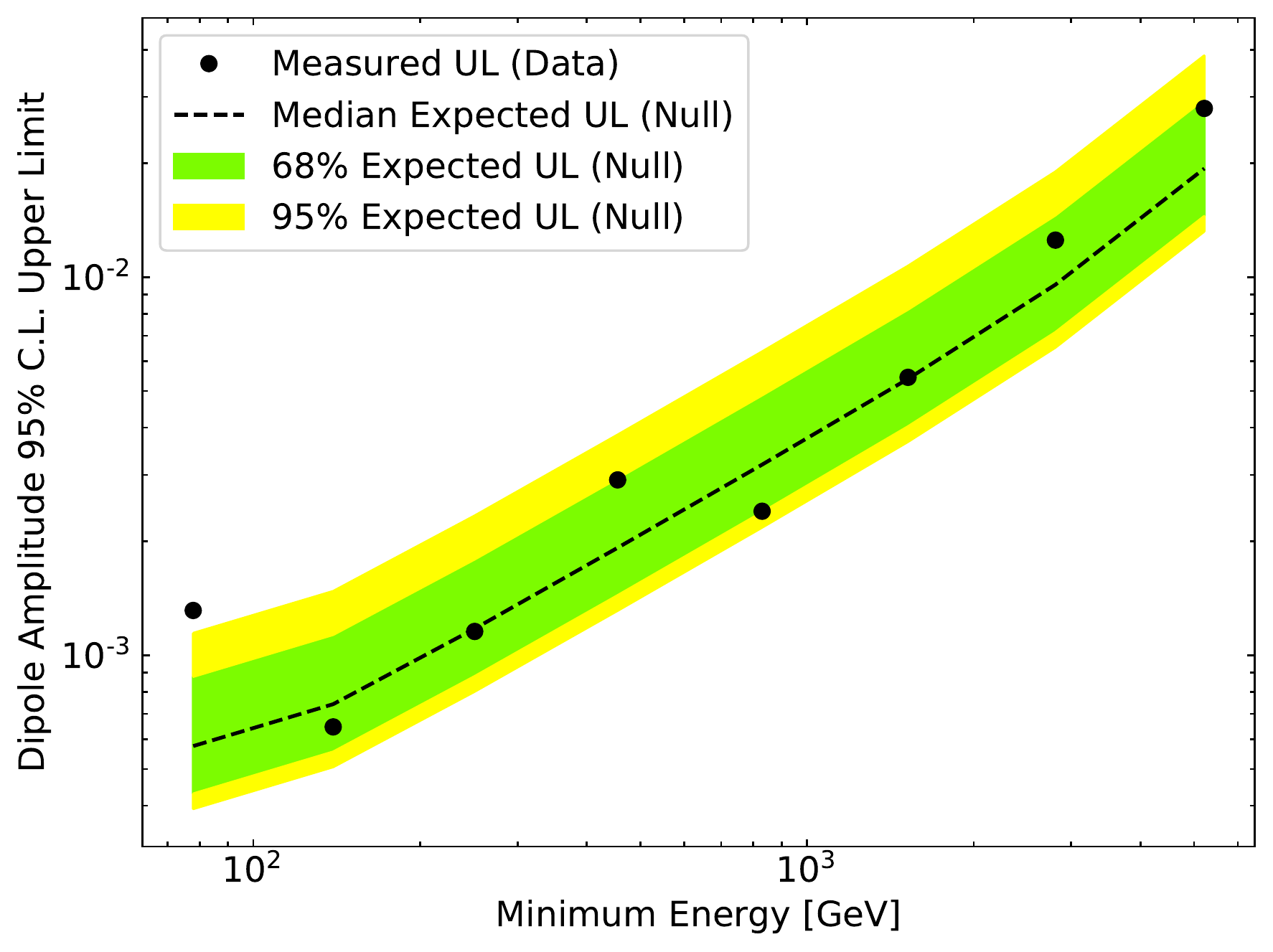}
	\caption{Top: Measured dipole amplitude, $\delta$, for each cumulative energy bin. The dipole amplitude is calculated directly from the measured dipole power ($C_1$). The error bars are the statistical uncertainty on the measured quantities and the colored bands represent the distribution of expected results under the null hypothesis, i.e., isotropic sky. Bottom: 95\% confidence level upper limits on the dipole amplitude for each cumulative energy bin. The dashed line represents the expected upper limits under the null hypothesis. The colored bands show the 68\% and 95\% containment of upper limits expected from an ensemble of measurements under the null hypothesis.}
	\label{fig:dipole_meas}
\end{figure}

Dipole amplitudes for each cumulative energy bin were measured using the dipole power ($C_1$) calculated for each bin and can be seen in Figure~\ref{fig:dipole_meas}. The dipole excess at E$>$78 GeV corresponds to the excess seen in the angular power spectrum in Figure~\ref{fig:power_spectrum}. The measured dipole amplitudes in the remaining energy intervals are all consistent with an isotropic sky. The exact amplitude and direction of the dipole excess are shown in Table~\ref{tbl:dipole}.

\begin{deluxetable}{ccccc}

\tabletypesize{\footnotesize}
\tablewidth{0pt}

\tablecaption{Spherical Harmonic Coefficients}

\tablehead{
	\colhead{Min. Energy [GeV]} & \colhead{$a_{1-1}$ [$10^{-3}$]} & \colhead{$a_{10}$ [$10^{-3}$]} & \colhead{$a_{11}$ [$10^{-3}$]} & \colhead{$\sigma_{stat}$ [$10^{-3}$]}
}

\startdata
78   & -0.39 & -0.61 & -0.57 & 0.28 \\
139  & -0.27 & -0.16 & -0.32 & 0.36 \\
251  & 0.37  & 0.72  & 0.26  & 0.56 \\
455  & 1.29  & 1.34  & -0.95 & 0.92 \\
830  & 0.83  & 0.65  & 0.90  & 1.53 \\
1522 & 2.33  & -2.13 & 2.47  & 2.58 \\
2810 & 1.95  & -1.87 & 8.68  & 4.58 \\
5218 & -4.78 & -8.40 & 17.66 & 9.29 \\
\enddata

\vspace{0.5cm}
\tablecomments{There is a maximum energy of $\sim$10 TeV.}
\label{tbl:sph_harm_coeff}

\end{deluxetable}

\begin{deluxetable}{cccccc}

\tabletypesize{\footnotesize}
\tablewidth{0pt}

\tablecaption{Observed amplitude and direction of the dipole excess and 95\% CL upper limit}

\tablehead{
	\colhead{Min. Energy [GeV]} & \colhead{$\delta_{obs}$ [$10^{-4}$]} & \colhead{RA [\textdegree]} & \colhead{Dec [\textdegree]} & \colhead{$\delta_{UL}^{95\%}$ [$10^{-3}$]}
}

\startdata
78   & 3.9 $\pm$ 1.5 & 215 $\pm$ 23 & $-$51 $\pm$ 21 & 1.3\\
\enddata

\vspace{0.5cm}
\label{tbl:dipole}

\end{deluxetable}

In Section~\ref{sec:systematics} we discuss the major sources of systematic uncertainty in the analysis that could lead to a dipole excess. We do not expect any of the systematics to create an excess at the level seen in our data. However, we cannot completely rule out the signal vs. systematic interpretation of the excess. We therefore computed upper limits on the total dipole amplitude, for all cumulative energy bins (including the excess) which can be seen in Figure~\ref{fig:dipole_meas}. The 95\% CL upper limits were calculated using the frequentist likelihood ratio approach used in~\citet{abdollahi_search_2017}. The upper limits on the observed dipole power are calculated after enforcing that $\hat{C_\ell}^{true}>0$ and then converted to upper limits on the dipole amplitude using Equation~\ref{eq:delta}. The 95\% CL upper limit on the dipole amplitude at a minimum energy of 78 GeV is $\delta_{UL} = 1.3\times 10^{-3}$. This calculated upper limit is considerably larger than the median expected upper limit of $5.8 \times 10^{-4}$ because it was calculated from an observed data point with an upward fluctuation.

\subsection{Systematics}\label{sec:systematics}
In the following subsections we describe the major sources of systematic uncertainty in the anisotropy measurement and the techniques we used to mitigate or quantify them in a data-driven way. We describe the ``east-west" effect seen in our data, the selection employed to reduce the tails of the PSF, and finally discuss the stability of the event rate and its effect on the results.

\subsubsection{East-West Effect}\label{sec:east_west}
\begin{figure}[h]
	\centering
    \includegraphics[width=\columnwidth]{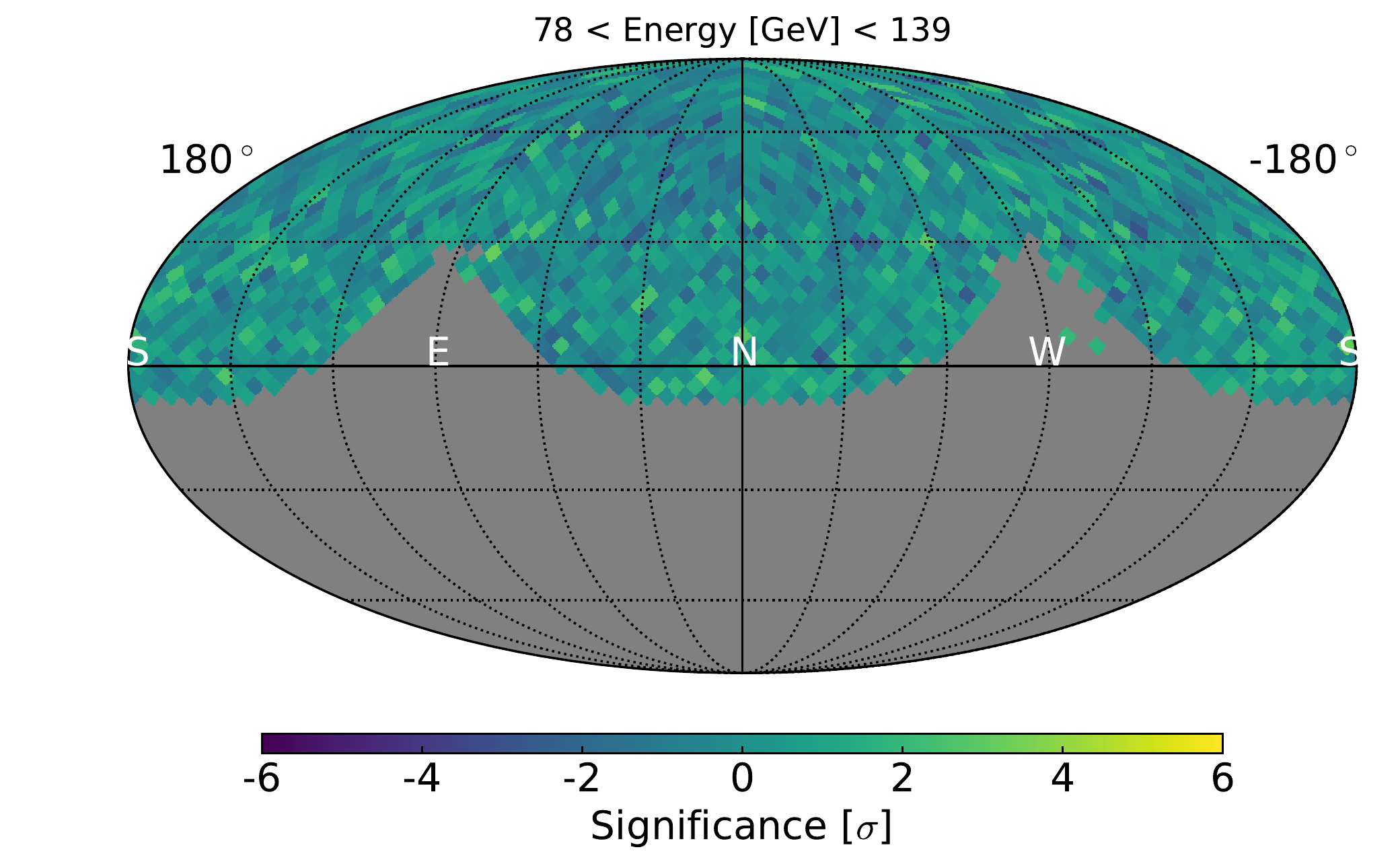}
    \includegraphics[width=\columnwidth]{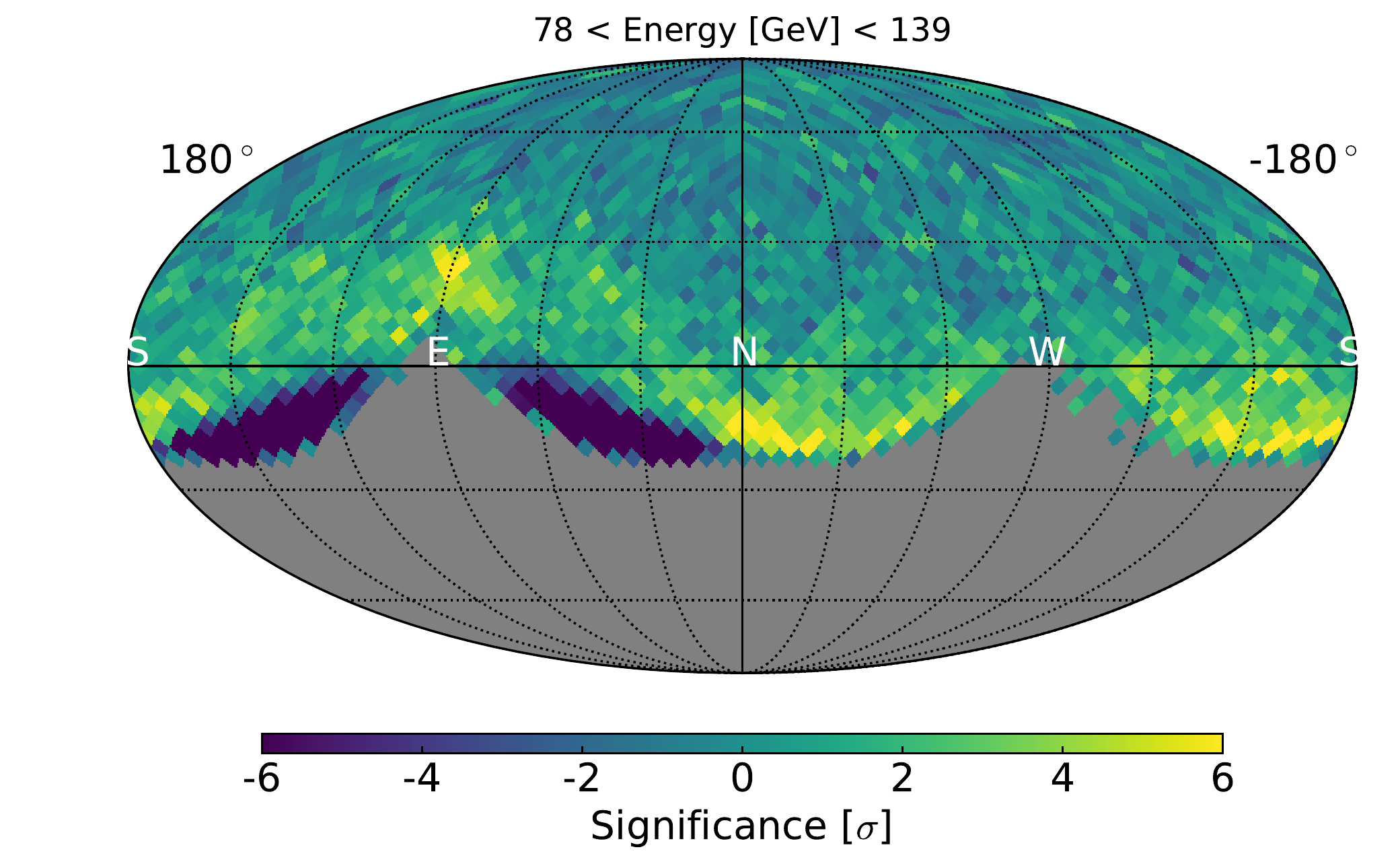}
	\caption{Sky maps in altitude-azimuth coordinates for events with a maximum off-axis angle cut of 45\textdegree{ }(top) and 60\textdegree{ }(bottom). Altitude is the complement of the zenith angle and the azimuth is measured from N=0\textdegree { }and increases towards the E. The ``east-west" effect is clearly visible in the data when a larger field of view is allowed.}
	\label{fig:altaz_maps}
\end{figure}

Although the cosmic rays in this data set are well above the vertical cutoff rigidity, geomagnetic effects are not completely negligible. Positively charged cosmic rays arriving from near the horizon from the east are ablocked by the Earth because their trajectories bend downward into the atmosphere. The famous ``east-west'' effect is visible in our data set if it is not accounted for. We impose energy-dependent off-axis angle (instrument theta) thresholds to mitigate this effect, which are similar to those used in the most recent {\it Fermi}-LAT $e^+/e^-$ anisotropy search~\citep{abdollahi_search_2017}. In the aforementioned study, the off-axis angle thresholds were optimized using simulations that estimated the geomagnetic influence by back-tracing events through a model of the geomagnetic field. However, this method relies on the accuracy of the energy estimation to properly trace the particle trajectories. The relatively poor energy resolution for protons in the LAT does not allow for accurate back-tracing to quantify the effect to the desired precision. Therefore, the selection was determined by analyzing the data in altitude-azimuth coordinates. The geomagnetic effects are maximal in this frame and the analysis remains unbiased to the equatorial anisotropy of interest. We created reference maps in differential energy bins for a range of maximum off-axis angle cuts in 5\textdegree{ }increments. We then analyzed significance maps in altitude-azimuth coordinates and performed a $\chi^2$ comparison between the significance distribution and a standard normal distribution, requiring $\chi^2_{red} \sim 1$. The analysis resulted in a maximum off-axis angle of 45\textdegree{ }in the lowest energy bin and 50\textdegree{ }for all others. Significance maps for the final selection and a selection where the east-west effect is visible in the lowest energy bin are shown in Figure~\ref{fig:altaz_maps}. The east-west effect is clearly visible in the lower panel where there is a $\gg$ 6$\sigma$ deficit of cosmic rays from the east. After reducing the field of view (top panel), there is no longer any visible anisotropy.

\subsubsection{Point Spread Function Tail}\label{sec:psf_tail}
\begin{figure}[h]
	\centering
  	\includegraphics[width=\columnwidth]{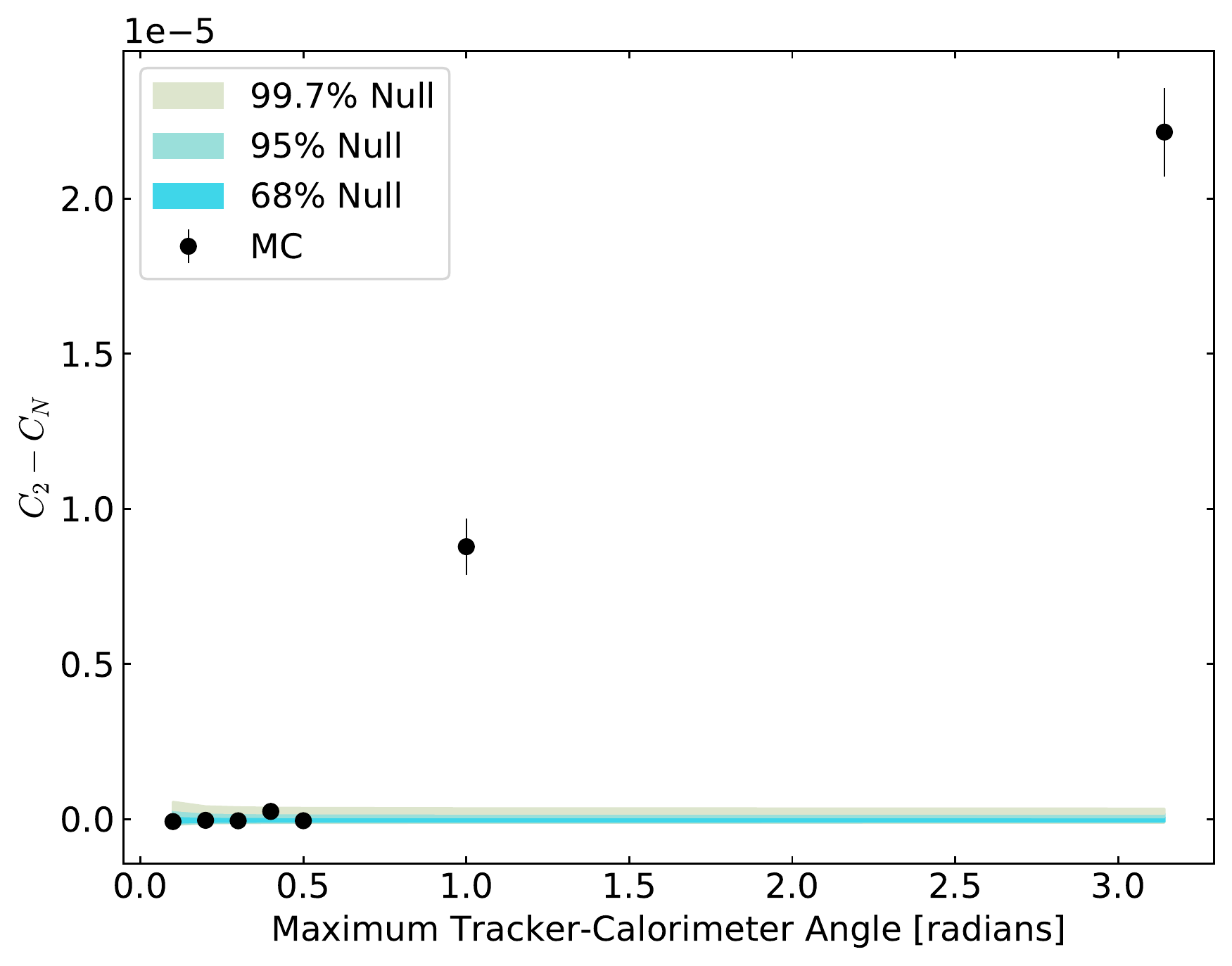}
    \includegraphics[width=\columnwidth]{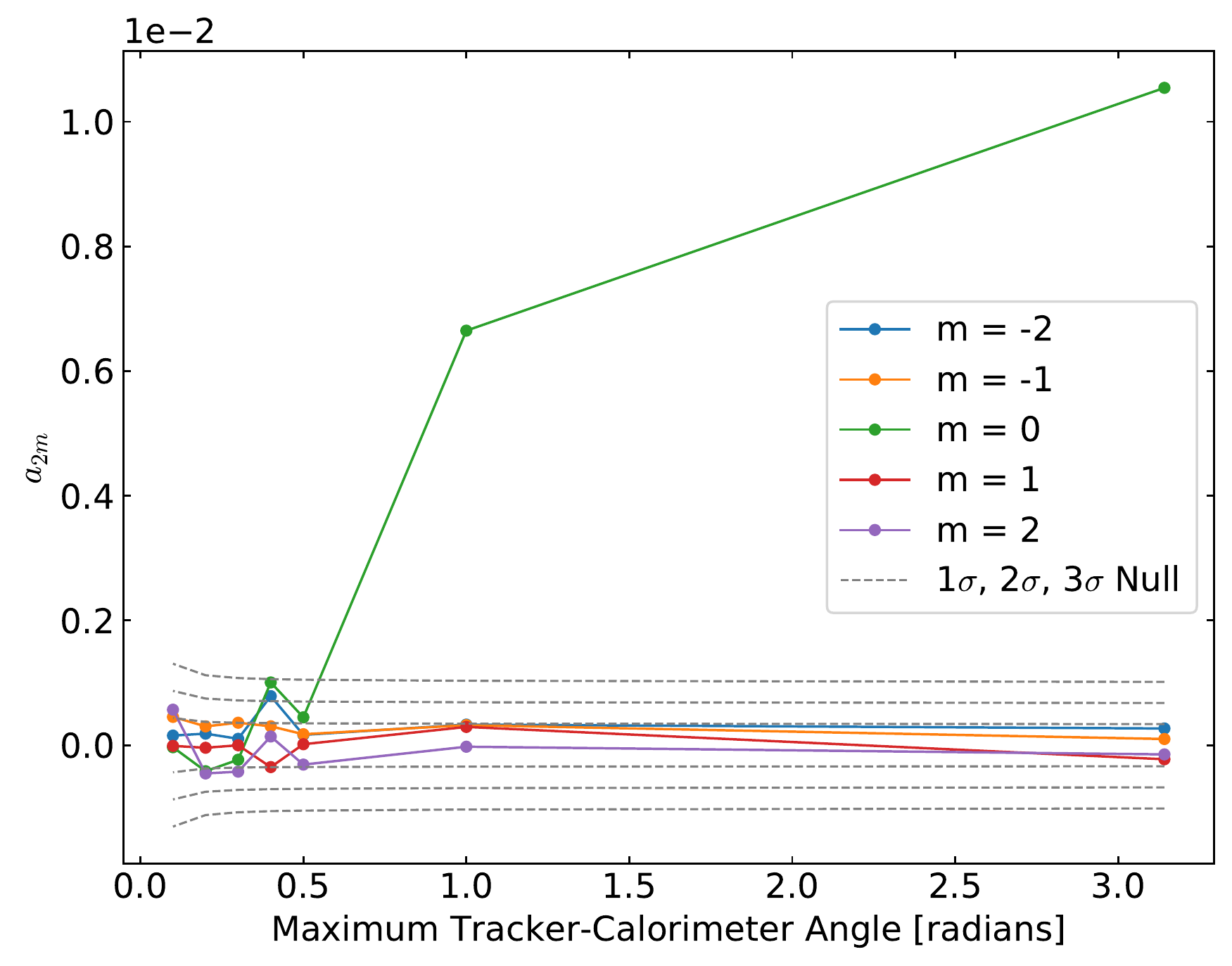}
  	\caption{Results of simulations of an 8-year observation with decreasing tracker-calorimeter angle thresholds. Top: The quadrupole power ($C_2$) is plotted vs. the maximum tracker-calorimeter angle used in the event selection. A significant quadrupole excess is detected at larger thresholds due to events from the PSF tail. Bottom: The $\ell$=2 coefficients of the spherical harmonic transform are plotted vs. the maximum tracker-calorimeter angle used in the event selection. All of the excess quadrupole power seen in the plot on the left is in the m=0 moment, i.e., $a_{20}$.}
  	 \label{fig:c2_vs_CalTrackAngle}
\end{figure}

Preliminary results of this analysis from 2017 measured a significant quadrupole excess~\citep{meehan_search_2017}. Further exploration indicated that the excess was caused by events from the tail of the PSF with large reconstruction errors. As stated in Section~\ref{sec:data_set}, the angular resolution for protons detected by the LAT is $\sim$0.01\textdegree. However, the tail of the distribution is non-negligible for an analysis at this level of sensitivity. The tail extends out to an angular error of 180\textdegree, which primarily consists of events that entered the bottom of the detector but were reconstructed as if they entered the top~\footnote{The LAT has no \emph{a priori} ability to tell the whether an event entered the top or bottom of the detector from timing alone.}. In this updated analysis we utilized an additional variable to remove events from the PSF tail. We compared independent direction measurements of each event from the tracker and calorimeter, and used the angle between them, hereafter tracker-calorimeter angle, as a proxy for the quality of the reconstruction. We created a detailed, Geant4-based simulation of our data set that includes realistic detector effects to accurately determine the tracker-calorimeter angle threshold necessary to reduce contamination from poorly reconstructed events. Figure~\ref{fig:c2_vs_CalTrackAngle} shows the quadrupole power vs. maximum tracker-calorimeter angle when thresholds were applied to this simulated data set and the rest of the anisotropy analysis was applied. There is a significant quadrupole excess for large maximum threshold and it is reduced for smaller thresholds, yielding the expected behavior. Additionally, the lower panel of Figure~\ref{fig:c2_vs_CalTrackAngle} shows that the entirety of the excess is in the $a_{20}$ coefficient of the multipole expansion. This is the moment in which exposure-related systematics are expected to exist. Recall that the LAT's exposure is primarily quadrupolar and aligned with the celestial poles (see Figure~\ref{fig:sky_maps}). Events that are reconstructed close to 180\textdegree{ }off of their true direction are likely to pile up at the equatorial poles, creating an $a_{20}$ excess.

\begin{figure}[h]
	\centering
  	\includegraphics[width=\columnwidth]{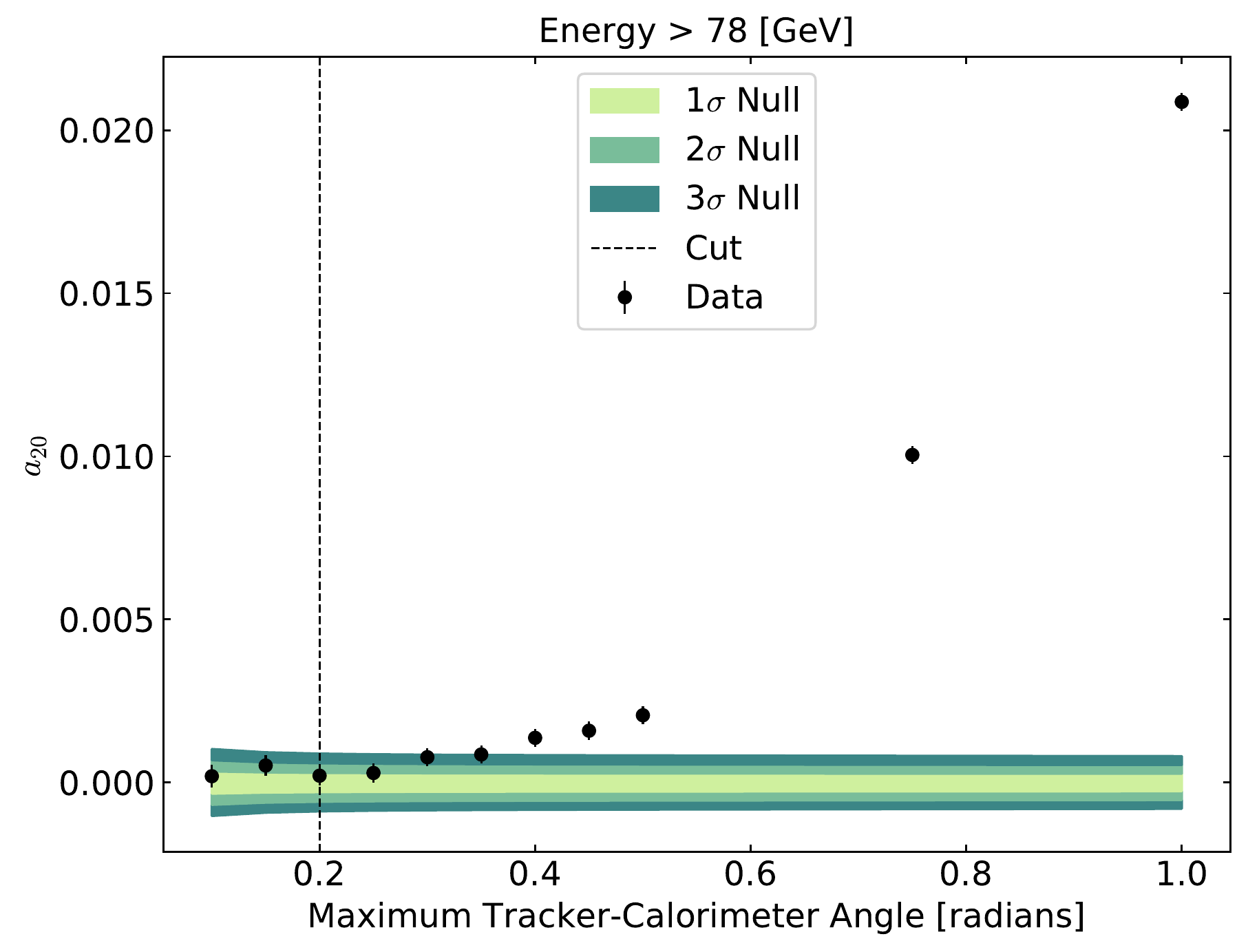}
  	\caption{Results of the scan over the tracker-calorimeter angle parameter using flight data. We scanned maximum tracker-calorimeter angle thresholds and ran the resulting data sets through the full anisotropy pipeline. To not bias the measurement of the other moments of the spherical harmonic analysis, we only utilized the $a_{20}$ in the scan, remaining blind to other the spherical harmonic parameters.}
  	 \label{fig:a20_vs_caltrackangle_4.50_4.75}
\end{figure}

While the simulation described above qualitatively explains the quadrupole, data/MC agreement was not precise enough to use the MC to determine an appropriate cut value. To tune the cut while staying unbiased to the parameters of interest, namely the dipole components of the analysis, we performed a parameter scan similar to that above, but only observed the $a_{20}$ moment of the spherical harmonic expansion. Figure ~\ref{fig:a20_vs_caltrackangle_4.50_4.75} shows the results of this parameter scan on flight data. The $a_{20}$ component is very significant for large maximum tracker-calorimeter angle thresholds and decreases as the threshold decreases. The final cut of 0.2 radians is indicated by the dashed line in the plot and the observed $a_{20}$ value is consistent with isotropy.

\subsubsection{Event Rate Stability}\label{sec:rate_stability}
The algorithm to construct the reference maps for the anisotropy search described in Section~\ref{sec:reference} uses the event rate averaged over each year of the data set to generate events. While the rate fluctuates on short time scales, the assumption is that it is stable over the entire year and also does not exhibit variation that is correlated with sky direction. Variation in the rate that is not properly accounted for in the reference map algorithm could over- or underestimate the exposure and result in a false-positive anisotropy. We tested both of these assumptions and describe our findings in the following section.

\begin{figure}[h]
	\centering
    \includegraphics[width=0.48\columnwidth]{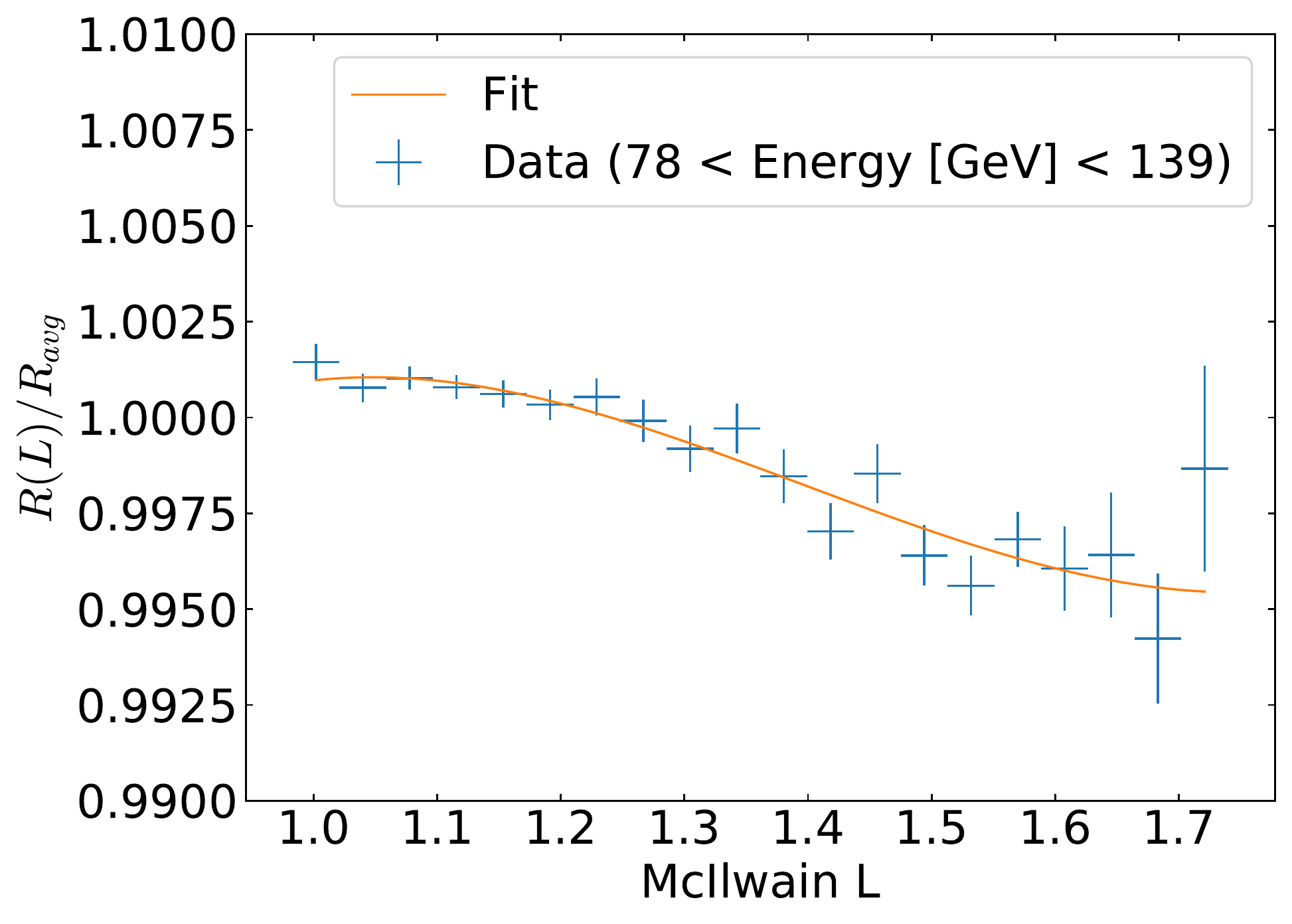}
    \includegraphics[width=0.48\columnwidth]{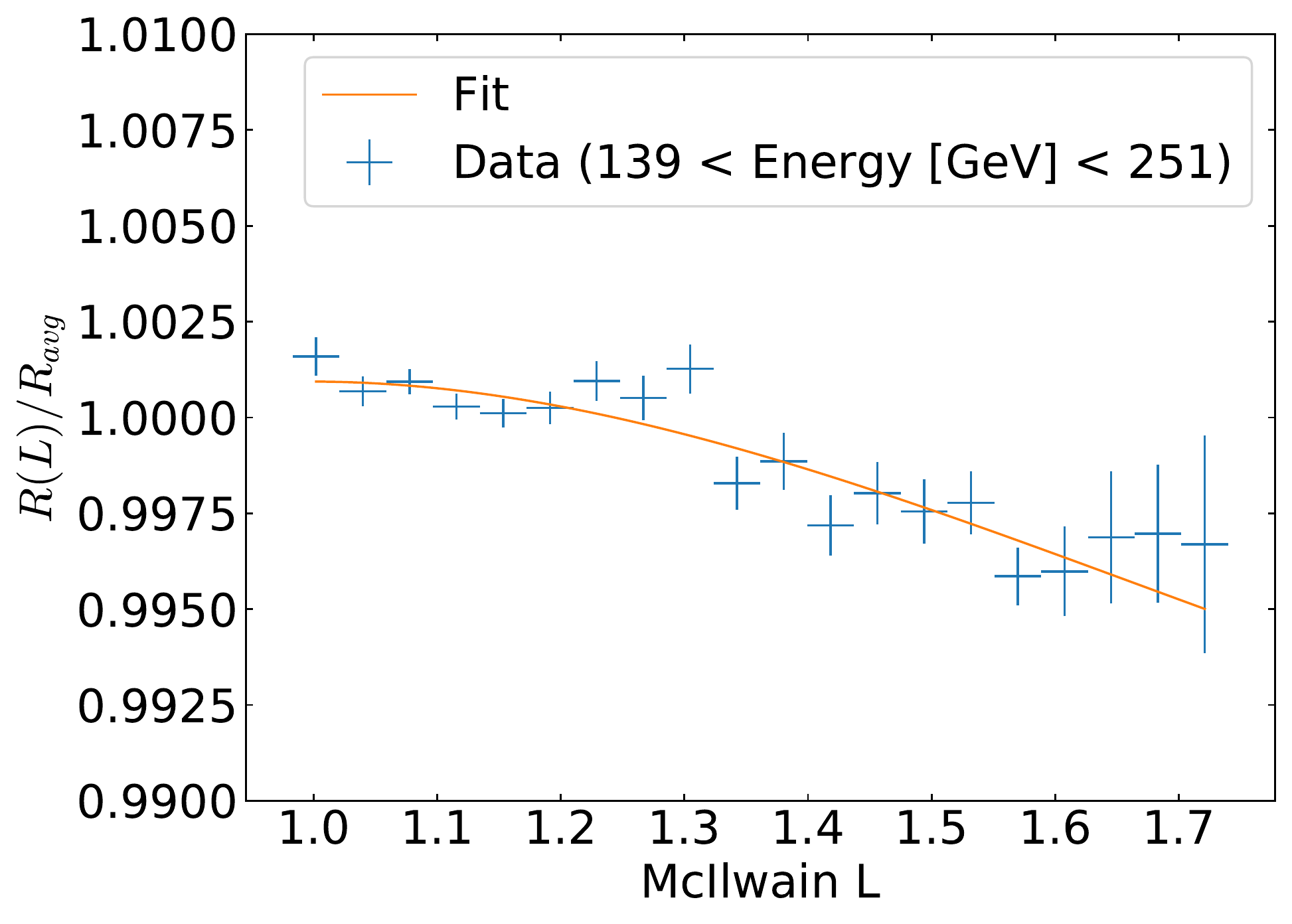}
    \includegraphics[width=0.48\columnwidth]{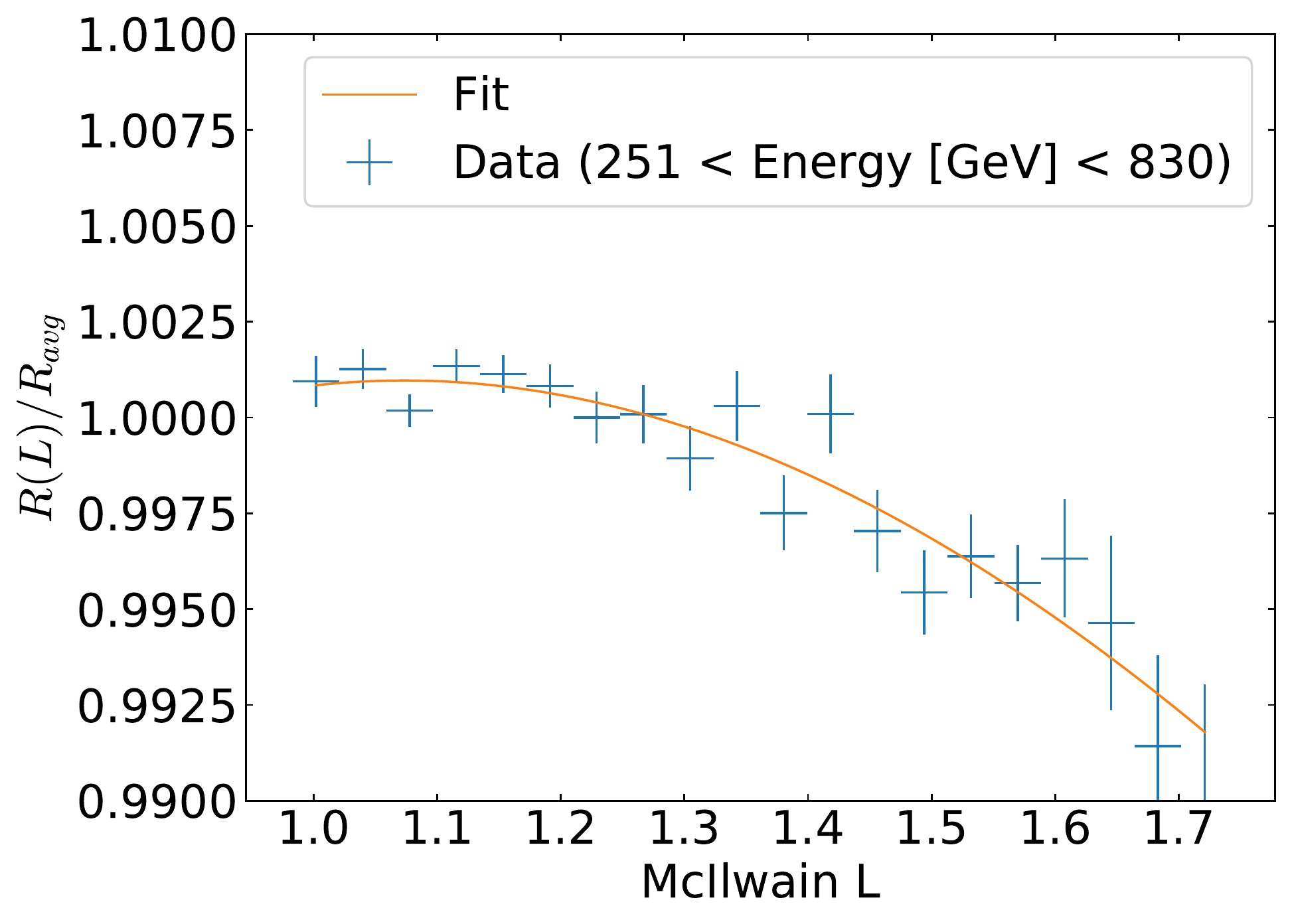}
    \includegraphics[width=0.48\columnwidth]{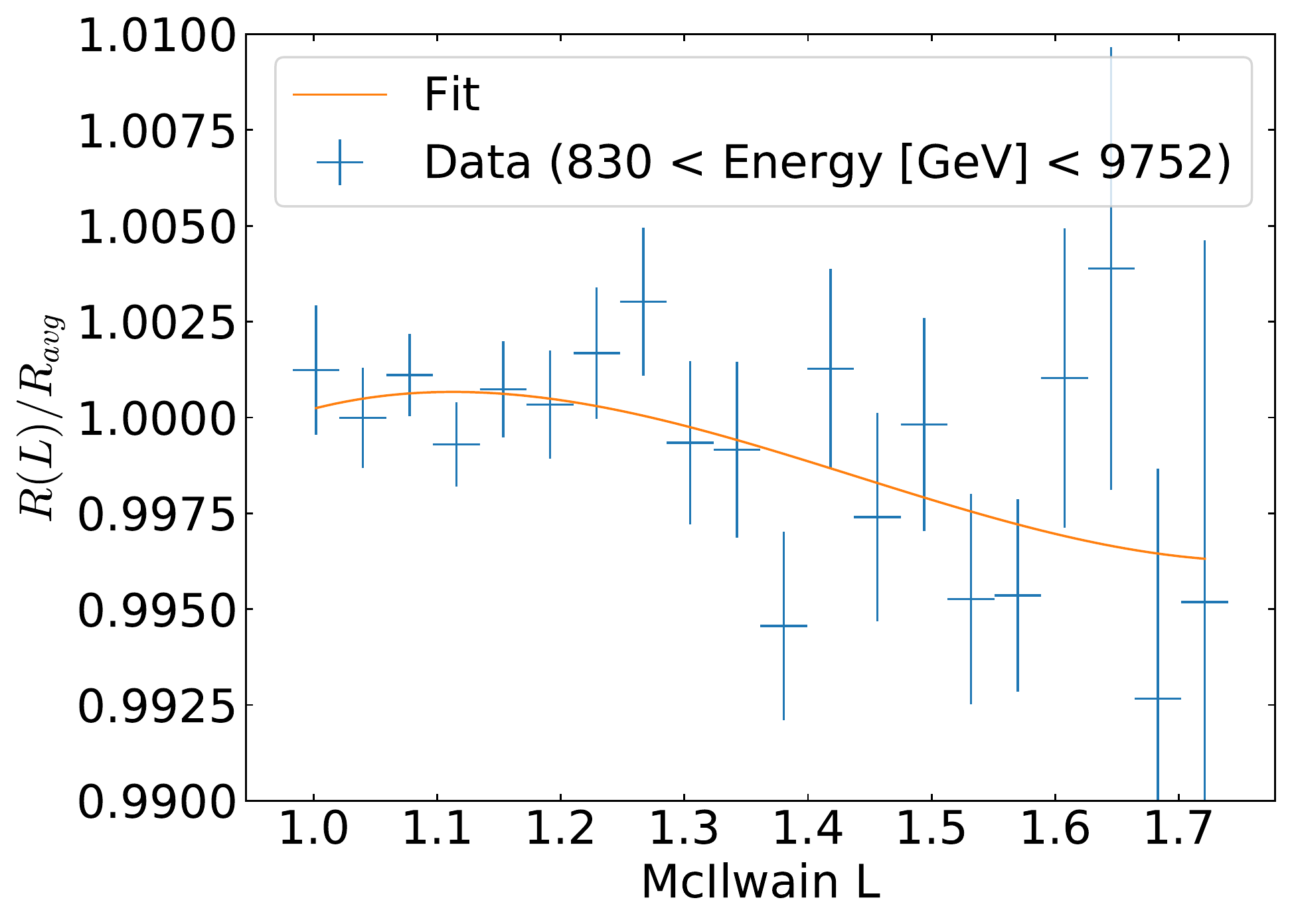}
  	\caption{Third-degree polynomial fits of the relative event rate in four energy bins. The eight analysis energy bins are combined to yield similar event statistics in each combined bin. The relative rate is the total eight year event rate in each McIlwain L bin divided by the average rate across all McIlwain L bins.}
\label{fig:rate_fits}
\end{figure}

\begin{figure}[h]
	\centering
    \includegraphics[width=\columnwidth]{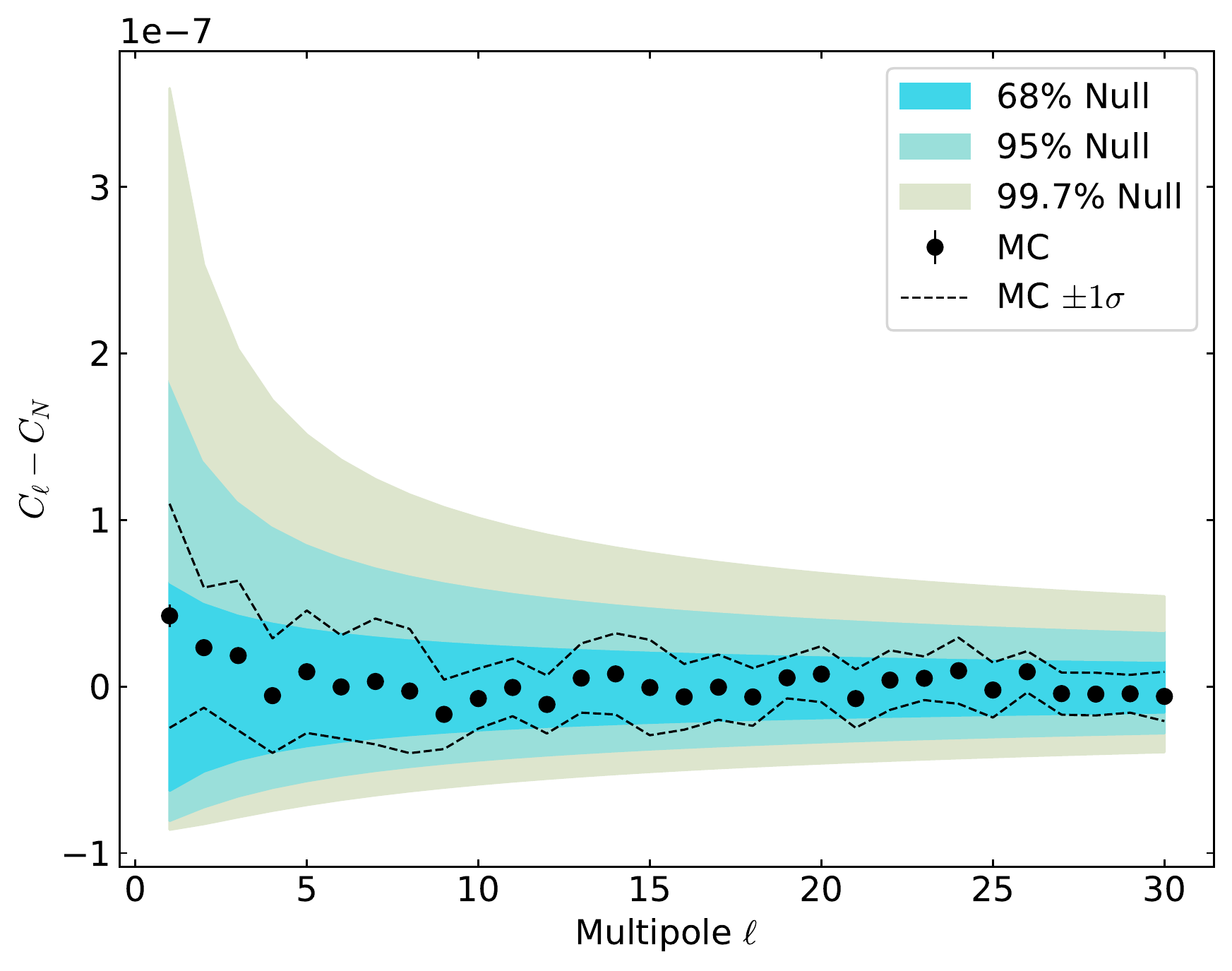}
  	\caption{Angular power spectra from an MC study of the McIlwain L-dependent event rate for event with a minimum energy of 78 GeV. Ten simulated data sets were generated using the McIlwain L-dependent rates in Figure~\ref{fig:rate_fits} and processed with the anisotropy pipeline. The data points are the mean angular power at each multipole from the ten realizations and the error bars, which are hardly visible, are the standard error on the mean. The dashed lines represent the 1$\sigma$ spread amongst the ten realizations. The colored bands represent the distribution of expected results under the null hypothesis, i.e., isotropic sky.}
\label{fig:mcilwain_mc}
\end{figure}

The geomagnetic cutoff varies continuously along the orbital path of {\it Fermi}. The increased rate of background cosmic rays in regions where the cutoff is lower could affect the efficiency of the various subsystems. To test this, we measured the event rate as a function of McIlwain L value of the LAT's location, which is directly correlated with the geomagnetic cutoff. The observed event rate can vary by as much as 1\% over the range of McIlwain L experienced by {\it Fermi}. To estimate the effect of this variation on the measured anisotropy we simulated ten data sets by first fitting the rate vs. McIlwain L curves in four energy bins and then generating events according to these curves~\footnote{The exact curves used to seed this MC study were fit to data using a slightly different event selection than the one described here. The shape and magnitude of the McIlwain L dependence in the MC study is statistically consistent with that in the current data set.}. Figure~\ref{fig:rate_fits} shows the observed event rate and third-degree polynomial fits for each of the four energy bins. As a full-circle test, we then processed the simulated data sets with the anisotropy search pipeline and measured the angular power spectrum for each realization. The results of this simulation are summarized in Figure~\ref{fig:mcilwain_mc}. The mean dipole power among the realizations was $\sim 1\sigma$ above the expectation under the null hypothesis. In other words, we expect a systematic $1\sigma$ dipole excess due to the McIlwain L-dependent event rate. We therefore conclude that this effect is not likely to create the observed dipole excess. In principle the distribution of dipole power that could be induced by this effect in an ensemble of experiments could have long tails. We leave the full characterization of this distribution to future work, as it would require significantly more realizations of the Monte Carlo simulation study described above.

\subsubsection{Raw Energy Threshold}\label{sec:raw_thresh}
As described in Section~\ref{sec:data_set}, the onboard filter utilizes a trigger that passes any event that deposits a minimum of 20 GeV of raw energy in the calorimeter to filter for high-energy particles. Radiation damage degrades the light yields of the calorimeter crystals over time. The raw energy used by the onboard filter is not calibrated for this decrease and the effective energy to pass the filter thus increases over time. This results in a time-dependent event rate for events with energies near the threshold. Without correcting for this effect, we observe a 0.5--1\%/year decrease in the event rate for events with energy in the range 78--139 GeV. To mitigate this effect, we imposed a minimum threshold of 21 GeV of calibrated energy deposited in the calorimeter, i.e., energy calibrated for the decreasing light yields. The 21 GeV threshold is above the effective threshold of the uncalibrated onboard filter energy. This mitigates most of the time-dependent effect, resulting in a total event rate that decreases by only $\sim 0.1$\%/year. We performed similar simulation studies to those described in Section~\ref{sec:rate_stability} to test the effect of this time dependence on the observed anisotropy. The study predicts that the systematic uncertainty in the anisotropy introduced by the monotonically decreasing rate is negligible compared to the statistical uncertainty in the data set and therefore will not affect the results.

\subsubsection{Stability of the Angular Distribution} \label{sec:angular_stability}
In addition to the event rate, the observed angular distribution of events is assumed to be stable on long time scales and with changing geographic location to estimate the reference sky maps. In principle, the time-dependent raw energy threshold described in Section~\ref{sec:raw_thresh} could create a time-dependent incidence angle distribution since the raw energy deposited in the calorimeter is a function of the event's path length. However, we do not measure any significant time dependence in this distribution. Additionally, we searched for McIlwain L dependence of the two-dimensional angular distribution of events. The distribution does not show any significant variation correlated with McIlwain L and is not expected to introduce any systematic uncertainties into the measurement.

\section{Conclusion}
We presented the results of the first search for cosmic-ray proton anisotropy using data from the {\it Fermi} Large Area Telescope. The 8-year data set is the largest single instrument, full-sky data set studying cosmic-ray anisotropy at these energies to date. As such, it provides the most stringent constraints on the declination dependence of the dipole anisotropy, which is not accessible by ground-based observatories.

Interpretation of the measured dipole excess is difficult due its marginal statistical significance and the complex systematics of the analysis. We discussed three potential sources of systematics and our method for quantifying or mitigating them: the ``east-west'' effect, poorly reconstructed events from the PSF tail, and the McIlwain L-dependent event rate due to the varying rate of background cosmic rays. Of these, only the last is expected to have a measurable effect on the dipole anisotropy. Our simulation study described in Section~\ref{sec:rate_stability} predicts a $1\sigma$ dipole excess due to this effect, but this does not fully explain the observed excess. Similar geomagnetic effects were seen in cosmic-ray anisotropy searches by the Alpha Magnetic Spectrometer (AMS-02) experiment, where a method was developed to correct for the systematic shift in the measured anisotropy~\citep{gebauer_measurement_2017, bindel_study_2017}. It may be possible to employ a similar correction in future anisotropy searches using LAT data. However, it is important to note that the orbit of AMS-02, which is onboard the International Space Station (ISS), has an inclination of 51.6\textdegree { }(cf. the LAT's orbital inclination of 25.6\textdegree) and travels through geographic locations with lower rigidity cutoffs than the LAT does. Geomagnetic effects are therefore expected to be less significant, in general, for the LAT anisotropy search. 

Residual contamination from other particles could also introduce systematic uncertainties into the measurement. Proton energies in this analysis were estimated by re-scaling the estimated gamma-ray energy to account for the missing portion of the hadronic shower in the calorimeter. This will also re-scale the accurately estimated energies of cosmic electrons and positrons (CREs) by a factor of $\sim3$, thereby introducing low-energy CRE contamination into the data set. Estimates from Geant4 Monte Carlo simulation place the contamination from CREs at less than 0.1\%. These CREs could be affected by the Heliospheric magnetic field, but it is difficult to quantify the effect on the anisotropy measurement. There is also residual contamination from Helium nuclei that is estimated to be less than 1\% (Section~\ref{sec:data_set}). As a cross-check, we performed the anisotropy analysis on a selection of Helium nuclei, which yielded a null result, i.e., consistent with isotropy. We therefore conclude that no systematic uncertainty should be introduced into the proton anistotropy measurement by residual, isotropic Helium in the data set.

The statistical excess of the dipole amplitude is in a difficult regime to make a strong statement about its interpretation. The measured dipole can be described in terms of its amplitude and two-dimensional direction in equatorial coordinates: ($\delta$, RA, Dec) = 3.9$\pm1.5 \times 10^{-4}$, 215\textdegree$\pm$23\textdegree, $-$51\textdegree$\pm$21\textdegree. Previous measurements of cosmic-ray anisotropy in the 100 GeV energy range by underground muon telescopes observed dipole amplitudes of $\sim2\times10^{-4}$ with maxima at right ascensions in the range $\varphi_0 \in [45^\circ,135^\circ]$~\citep{swinson_corrected_1985, hall_gaussian_1999}, where $\varphi_0$ is the phase of the one-dimensional dipole fit typically performed by ground-based experiments. The phase of the TeV anisotropy described in Section~\ref{sec:intro} is similar to that measured by underground muon telescopes and is typically in the range $\varphi_0 \in [30^\circ,50^\circ]$. We note that the direction of the dipole measured in this analysis is in tension with the measurements by muon telescopes from decades ago, but stress the many differences between the analyses. These telescopes typically scanned a small patch of overhead sky and recorded the daily sidereal variation in the counting rate as a function of right ascension, while the $Fermi$-LAT analysis measures the all-sky, two-dimensional anisotropy. Additionally, ground-based experiments have limited charge resolution compared to the LAT and measure the all-particle anisotropy, rather than the measurement of protons only as presented here. The myriad differences between these measurements are important to understand to elucidate the origin of the anisotropy. Given the lack of a definitive signal, we set upper limits on the total dipole amplitude: the 95\% CL upper limit on the dipole amplitude is $\delta_{UL}=1.3\times 10^{-3}$ for protons with a minimum energy of 78 GeV and $\delta_{UL}=1.2 \times 10^{-3}$ for protons with a minimum energy of 251 GeV. Recently, a nearly all-sky measurement of cosmic-ray anisotropy was performed at $\sim10$ TeV by combining data sets from the IceCube and HAWC experiments~\citep{abeysekara_all-sky_2018}, resulting in the least-biased measurement of the anisotropy to date. However, this measurement is still insensitive to the declination component of the dipole due to the limitations of ground-based measurements mentioned in Section~\ref{sec:intro}. Our upper limits are therefore the most constraining on the declination dependence of the dipole by any experiment.

\acknowledgments
The \textit{Fermi} LAT Collaboration acknowledges generous ongoing support
from a number of agencies and institutes that have supported both the
development and the operation of the LAT as well as scientific data analysis.
These include the National Aeronautics and Space Administration and the
Department of Energy in the United States, the Commissariat \`a l'Energie Atomique
and the Centre National de la Recherche Scientifique / Institut National de Physique
Nucl\'eaire et de Physique des Particules in France, the Agenzia Spaziale Italiana
and the Istituto Nazionale di Fisica Nucleare in Italy, the Ministry of Education,
Culture, Sports, Science and Technology (MEXT), High Energy Accelerator Research
Organization (KEK) and Japan Aerospace Exploration Agency (JAXA) in Japan, and
the K.~A.~Wallenberg Foundation, the Swedish Research Council and the
Swedish National Space Board in Sweden.
 
Additional support for science analysis during the operations phase is gratefully
acknowledged from the Istituto Nazionale di Astrofisica in Italy and the Centre
National d'\'Etudes Spatiales in France. This work performed in part under DOE
Contract DE-AC02-76SF00515.

We would like to thank the INFN GRID Data Centers of Pisa, Trieste and CNAF-Bologna, the DOE SLAC National Accelerator Laboratory Computing Division and the CNRS/IN2P3 Computing Center (CC-IN2P3 - Lyon/Villeurbanne) in partnership with CEA/DSM/Irfu for their strong support in performing the massive simulations necessary for this work.

\bibliography{Zotero.bib}
\end{document}